\documentclass{llncs}

\usepackage{amsthm}

\usepackage{lipsum}
\usepackage{amsfonts}
\usepackage{graphicx}
\usepackage{epstopdf}
\usepackage{algorithmic}

\title{On Ring Learning with Errors over the Tensor Product of Number Fields\thanks{This work is partially funded by the Agencia Estatal de Investigaci\'{o}n (Spain) and the European Regional Development Fund (ERDF) under projects WINTER (TEC2016-76409-C2-2-R), by the Xunta de Galicia and the European Union (European Regional Development Fund - ERDF) under projects Agrupación Estrat\'{e}xica Consolidada de Galicia accreditation 2016-2019 and Red Tem\'{a}tica RedTEIC 2017-2018, and by the EU H2020 Programme under project WITDOM (project no. 644371).}}

\author{Alberto~Pedrouzo-Ulloa\inst{1}, Juan~Ram\'on~Troncoso-Pastoriza\inst{2} and~Fernando~P\'erez-Gonz\'alez\inst{1}}

\institute{Signal Theory and Communications Department, University of Vigo, Vigo, Spain \\
\email{\{apedrouzo,fperez\}@gts.uvigo.es} \and Laboratory for Communications and Applications 1, \'{E}cole Polytechnique F\'{e}d\'{e}rale de Lausanne, Lausanne, Switzerland \\\email{juan.troncoso-pastoriza@epfl.ch}}

\usepackage{graphicx}
\usepackage{color}
\usepackage{placeins}
\usepackage{float}
\usepackage{tabularx,colortbl}

\newtheorem{thmRLWE}{Theorem}
\newtheorem{lem}{Lemma}
\newtheorem{defin}{Definition}

\usepackage{amsmath}
\usepackage{amsfonts}
\usepackage{subfigure}
\usepackage{bm}
\usepackage{url}
\usepackage{multirow}

\usepackage{amssymb}
\usepackage{upgreek}

\begin{document}

\maketitle

\begin{abstract}
The ``Ring Learning with Errors'' (RLWE) problem was formulated as a variant of the ``Learning with Errors'' (LWE) problem, with the purpose of taking advantage of an additional algebraic structure in the underlying considered lattices; this enables improvements on the efficiency and cipher expansion on those cryptographic applications which were previously based on the LWE problem. In Eurocrypt 2010, Lyubashevsky \emph{et al.} introduced this hardness problem and showed its relation to some known hardness problems over lattices with a special structure. In this work, we generalize these results and the problems presented by Lyubashevsky \emph{et al.} to the more general case of multivariate rings, highlighting the main differences with respect to the security proof for the RLWE counterpart. This hardness problem is denoted as ``Multivariate Ring Learning with Errors'' ($m$-RLWE or multivariate RLWE) and we show its relation to hardness problems over the tensor product of ideal lattices. Additionally, the $m$-RLWE problem is more adequate than its univariate version for cryptographic applications dealing with multidimensional structures.
\end{abstract}


\section{Introduction}
\label{sec:intro}

In recent years, a high number of cryptographic schemes and applications have been proposed based on the LWE (Learning with Errors) problem. However, in spite of the versatility of this hardness assumption for developing cryptographic primitives, the main drawback of the cryptosystems whose security is based on LWE is their efficiency.
Actually, several schemes that allow to perform an unbounded number of encrypted operations (Fully Homomorphic Encryption Schemes, FHE) have been devised, but the needed size of the keys and the required computation times are still too high for practical applications.

In order to alleviate this issue, an algebraic version of the LWE problem was proposed by Lyubashevsky \emph{et al.} \cite{LPR13,LPR10}. This hardness assumption, called ring-LWE, is based on worst-case problems on ideal lattices instead of general lattices. Although the use of lattices with an additional algebraic structure could allow for the existence of better attacks, nowadays there are no known attacks to RLWE that get a substantial advantage with respect to attacks to LWE.\footnote{In \cite{ABD16}, Albrecht \emph{et al.} take advantage of the presence of a subfield in the considered number field which allows them to deal with an easier lattice problem. However, while this technique allows to have an attack for the overstretched NTRU problem, the RLWE problem is not affected.}

Hence, the RLWE problem and the analysis of its security reductions to hardness problems on ideal lattices have enabled the introduction of new cryptographic applications: Brakerski \emph{et al.}~\cite{BV11b,BV11aJ} proposed several versions of FHE cryptosystems, varying from leveled FHE schemes to the most recent scale-invariant versions \cite{FV12,Brakerski12,BLLN13}.

Practical applications from Secure Signal Processing (SSP) have made extensive use of homomorphic encryption~\cite{BPB09a}, and especially additive schemes like Paillier's \cite{Pa99}. However, the Paillier cryptosystem has several drawbacks for practical implementations, being its two main problems the very high cipher expansion and the inability to perform multiplications between two encrypted messages.

In order to resolve the first drawback, packing and unpacking steps were introduced in \cite{TKCL07,BPB10}; for the second drawback, several recent works resort to Somewhat Homomorphic Encryption (SHE) schemes \cite{TGP13} to enable simultaneous use of fully encrypted signals. While SHE schemes only allow for a limited number of encrypted operations, they are more efficient than their Fully Homomorphic counterparts. Therefore, if the number of operations that have to be performed under encryption is known beforehand (this is usually true in many practical applications), the use of SHE schemes increases the efficiency of the solution.

Nevertheless, when working with multidimensional signals, both Paillier cryptosystem and RLWE based cryptosystems present a very high cipher expansion (even after incorporating packing and unpacking techniques). In this context, the authors \cite{PTP15} introduced some example cryptosystems based on a variant of the RLWE problem called $m$-RLWE (multivariate Ring Learning with Errors) that extends RLWE from the univariate case to the multivariate one. These cryptosystems can be defined by extending to the multivariate case the most typical RLWE based cryptosystems. They bring about clear advantages in terms of efficiency and size of the underlying lattice when working with multidimensional signals, and they allow for packing several signals in only one ciphertext. 
It is also important to note that some of the contributions of \cite{PTP15} can be adapted to work with RLWE based cryptosystems considering the tensorial decomposition in ``coprime'' cyclotomic fields shown in the work of Lyubashevsky \emph{et al.} \cite{LPR13b}. This approach only requires to have enough space inside the polynomials in order to properly store the result of linear convolutions.

However, there are several applications that cannot be easily adapted to the RLWE case, like those presented in \cite{PTP16,PTPwifs}, because a particular modular function is needed to enable several usual operations belonging to the field of Signal Processing. In addition, having the same modular function on several variables can be a requirement in some cases, and this is not considered in the work of Lyubashevsky \emph{et al.} and can only be tackled by resorting to the $m$-RLWE problem.

While several comparisons between $m$-RLWE and RLWE have been presented considering basis-reduction attacks \cite{BKZ2,MR09} and decoding attacks as described in \cite{LP11}, a reduction from hardness problems on lattices and a complete security proof have not been provided yet; this is the main contribution of this work.

The main objective of this work is to adapt and generalize the techniques of Lyubashevsky \emph{et al.} \cite{LPR13} for the RLWE problem and achieve a reduction of the $m$-RLWE from hardness problems over ideal lattices, hence giving some new insights into its hardness. For the sake of completeness, we present a generalized version of the multivariate RLWE problem (introduced on \cite{PTP15}) which is not limited to only work with cyclotomic modular functions of degree power of two, hence it is possible to have any type of cyclotomic polynomial as modular function.

\subsection{Motivation and Contributions}
The ring structure of the RLWE problem allows for defining several cryptographic primitives as for example homomorphic cryptography; providing a ring homomorphism which enables both addition and multiplication of ciphertexts. Although we can consider different types of rings, for practical purposes the most used are those polynomial rings where the modular function is a cyclotomic polynomial with the form $1 + z^n$ being $n$ a power of two.

On the one hand, we can find an efficient implementation of the polynomial operations resorting to radix algorithms of the NTT (Number Theoretic Transforms). For example one of the most efficient libraries for homomorphic cryptography~\cite{ABGGKL16} deals with highly efficient NTT transforms to perform the different polynomial operations. On the other hand, polynomial operations over the previous ring correspond to basic blocks used in different practical applications belonging to Computer Vision and Signal Processing~\cite{PTP16}, covering linear convolutions, filterings, linear transforms and many more.

When working with multidimensional structures, as for example videos or images~\cite{PTP15,PTPwifs}, the use of the isomorphism with ``prime'' cyclotomic rings (see~\cite{LPR13,LPR10}) is not valid, because the use of several modular functions with the same form (for example $1 + z^n$) is required. This is the context where the authors of \cite{PTP15} present what they called $m$-RLWE problem as a means to easily deal with encrypted multidimensional structures.

Hence, our main purpose in this work is to provide a security reduction from hard lattice problems to the $m$-RLWE problem. Additionally, we generalize the $m$-RLWE variant introduced in~\cite{PTP15}, considering any type of cyclotomic polynomial as a modular function (not being restricted to the $1 + z^n$ case).

Our proof follows the techniques introduced by Lyubashevsky \emph{et al.}~\cite{LPR13,LPR10}, adapting and correcting their proof in order to deal with the new inconveniences that this tensor case introduces. For example, the considered number field tensor is not even a field (considering $(\mathbb{Q}[x,y] \bmod{1 + x^n}) \bmod{1 + y^n}$, the polynomial $x + y$ does not have inverse). We discuss and show how the different peculiarities of the tensor case can be tackled.

Therefore, our first step is to justify that the main properties which are required by the techniques from Lyubashevsky \emph{et al.} are preserved: a) we show that the ring homomorphism between the finite field tensor and the subspace $H_{(T)}$ exists (even though the finite field tensor is not a field); and we define the Gaussian measures over this tensor space, b) we explain the structure of the automorphisms which can be used in this tensor case and how to address and work with them; and c) we explain the use of the CRT (Chinese Remainder Theorem) and its effects over the corresponding automorphisms.

Additionally, we carefully readapt and revise the tools introduced by Lyubashevsky \emph{et al.} to the new properties; and those details (for example those which are involved on the proof of the results presented in Section~\ref{sec:mrlwe}) which need further treatment or corrections are consequently explained (see Appendices~\ref{sec:prelim},\ref{sec:proofhardsearchLWE} and \ref{sec:proofs4everyting}).

\subsection{Structure and notation}
\label{sec:not}
The structure of the paper is as follows: Section~\ref{specialproperties} extends some properties of cyclotomic number fields to the tensor product case.
Section~\ref{sec:mrlwe} introduces the $m$-RLWE problem together with the main theorem and the necessary definitions. Finally, Section~\ref{sec:proofsketch} presents the security reductions for $m$-RLWE along with the involved theorems, sketching their proof. Appendix~\ref{sec:prelim} revisits the necessary concepts of algebraic number theory and lattices when they are extended to the tensor of number fields, while Appendices~\ref{sec:proofhardsearchLWE} and \ref{sec:proofs4everyting} present the lemmas and proofs for the main reductions.

We denote matrices and vectors with uppercase and lowercase letters, respectively; $\langle \bm{a}, \bm{b} \rangle$ represents the scalar product between two vectors $\bm{a}$ and $\bm{b}$. For a vector $\bm{x}\in \mathbb{C}^n$ we define its $l_p$ norm as ${||\bm{x}||}_p = { \left( \sum_{i \in [ n ]} {|x_i|}^p \right) }^{1/p}$, where $1 \leq p < \infty$ with $p \in \mathbb{R}$, and ${||\bm{x}||}_{\infty} = \mbox{max}_{i \in [ n ]}|x_i|$. If $p$ is omitted, we consider the Euclidean norm.

The set $[n]$ is defined as $\{1, 2, \ldots, n\}$. We also work with some additional operators as the tensor product $\bigotimes$ and the direct sum $\bigoplus$. When dealing with number fields (or the corresponding ring of integers), as the tensor product is always defined over the rational numbers (integer numbers) we ignore the subscript if there is no ambiguity.

\section{Properties of the Tensor Product of Cyclotomic Number Fields}
\label{specialproperties}

For the sake of completeness, we discuss why the embeddings, automorphisms and even the Chinese Remainder Theorem (CRT, see Appendix~\ref{sec:CRT}) can be perfectly defined over the tensor of cyclotomic fields and the corresponding tensor ring of integers. 
Although the three previous concepts are interrelated, we separately explain their existence in the following sections.

The notation and tools used throughout this discussion are defined in Appendix~\ref{sec:prelim}, which introduces the main concepts needed to obtain the security proofs of the multivariate extension of the RLWE problem, by extending several of the concepts presented in \cite{LPR13} to our more general case. We refer to this appendix when needed, but we encourage the reader to go over it before reading this section.

\subsection{Embeddings}
\label{specialembedding}
We can work with the embedding over the space $H$ (see Appendix~\ref{sec:Ht}) of any type of cyclotomic field. Of course, as we can decompose a cyclotomic field in the tensor of power prime cyclotomic fields, it is easily shown that for that particular case of tensor of cyclotomic fields the embedding exists.

However, in our more general case this relation with cyclotomic fields does not necessarily hold, so we can not justify the existence of the tensor embedding by solely resorting to the existence of the embedding in an isomorphic cyclotomic field.

We can see that the embedding of a cyclotomic field (respectively, its corresponding ring of integers or the corresponding reduction modulo $q$) is equivalent to an invertible linear transformation from $\mathbb{Q}^{\phi(m_i)}$ (respectively, $\mathbb{Z}^{\phi(m_i)}$ or $\mathbb{Z}_q^{\phi(m_i)}$) to the corresponding subspace $H_i \subseteq \mathbb{C}^{n_i}$, where $n_i = \phi(m_i)$ (see Appendix \ref{sec:prelim}).

Now, there are two properties of Kronecker products that allow us to justify the existence of the embeddings. The first one is that $\mbox{det}(\bm{A} \bigotimes \bm{B}) = \mbox{det}(\bm{B} \bigotimes \bm{A})$ $= {(\mbox{det}(\bm{A}))}^n {(\mbox{det}(\bm{B}))}^m $ where $\bm{A}$ and $\bm{B}$ are square matrices of size $n \times n$ and $m \times m$, respectively. This property states that $\bm{A} \bigotimes \bm{B}$ is non singular (and therefore invertible) if and only if $\bm{A}$ and $\bm{B}$ are non singular. The second one is that ${\left(\bm{A} \bigotimes \bm{B} \right)}^{-1} = \bm{A}^{-1} \bigotimes \bm{B}^{-1}$, which defines this inverse. For more details about the different properties of the Kronecker product we refer the reader to \cite{HJ91}.

Additionally, we can see that our embedding can be defined as the Kronecker product of different invertible linear transformations that correspond to the different embeddings for each cyclotomic field. Hence, resorting to the properties of the Kronecker product we can see that there exists the corresponding tensor embedding between the tensor of cyclotomic fields and the subspace $H_{(T)} = \bigotimes_{i \in \left[ l \right]} H_i$ (see Appendix~\ref{sec:Ht}).

\subsection{Automorphisms and Linear Representation Theory}
In order to justify the structure and behaviour of the new automorphisms we resort to the theory of Linear Representations \cite{Serre77}. First, we introduce the main concepts needed from this theory, and afterwards, we detail the different automorphisms that we can find.

In general, we consider $V$ as a vector space of dimension $d$ over $\mathbb{C}$ and we define $\mbox{GL}(V)$ as the group composed of all the isomorphisms of $V$ onto itself.
An element $a$ belonging to $\mbox{GL}(V)$ can be seen as a linear mapping from $V$ to $V$ and we denote its inverse as $a^{-1}$. Analogously, we could think of each linear mapping as an invertible square matrix $A$ of size $d \times d$ whose coefficients are complex numbers. Hence, we can see that $\mbox{GL}(V)$ is composed of all the different invertible square matrices of order $d$.

Now, if we consider a finite group $G$, we define a linear representation of $G$ in $V$ as a homomorphism $\rho$ from $G$ to $\mbox{GL}(V)$. Considering that the group $G$ has the composition operation $(r, s) \rightarrow rs$ for $r,s \in G$, we have the following property:
\begin{equation*}
\rho(rs) = \rho(r)\rho(s),
\end{equation*}
where $\rho(r)\rho(s)$ represents the matrix multiplication operation between the two associated matrices to $r$ and $s$, respectively. Two important properties are that when $1 \in G$, this implies $\rho(1) = 1$ and $\rho(s^{-1}) = {\rho(s)}^{-1}$. Commonly, we consider $V$ as a representation space (or simply a representation) of $G$.

Now, we can particularize the previous results to our specific case, for $W = \mathbb{Q}\left( \varsigma_{m_i} \right) \subset \mathbb{C}$ (see Appendix \ref{sec:algnumbertheory}). If we consider $G = \mathbb{Z}_{m_i}^*$ and as the composition operation we consider the product operation between units of $\mathbb{Z}_{m_i}$, we have the following linear representation $\rho_i: \mathbb{Z}_{m_i}^* \rightarrow \mbox{GL}(\mathbb{Q}(\varsigma_{m_i}))$ where $\rho_i(\mathbb{Z}_{m_i}^*) \subseteq \mbox{GL}(\mathbb{Q}(\varsigma_{m_i}))$ is composed of the different automorphisms $\uptau_k = \rho_i(k)$ for $k \in \mathbb{Z}_{m_i}^*$ such that $\uptau_k(\varsigma_{m_i}) = \varsigma_{m_i}^k$, hence having $\mathbb{Q}(\varsigma_{m_i})$ as a representation of $\mathbb{Z}_{m_i}^*$. It is important to note that the effect of the automorphism $\uptau_k$ over the embedding is a rotation of the coordinates of the subspace $H_i$, that is, $\sigma_{i}(\uptau_k(\varsigma_{m_i})) = \sigma_{ik}(\varsigma_{m_i})$, being $i \in \mathbb{Z}_{m_i}^*$.

Of course, the linear representation preserves the linear structure and, in this case, as we have a commutative group $\mathbb{Z}_{m_i}^*$, there exists an equivalent representation such that each square matrix associated to each particular automorphism can be decomposed as a direct sum of $n$ irreducible representations $\bigoplus_{j \in \left[ n_i \right]} V_j$ (i.e., each irreducible representation for which the only decomposition is the trivial one $V_j = 0 \oplus V_j$). This implies that there exists an isomorphic domain where we can represent all the elements of $K_i$ in such a way that each different representation (different automorphism of $K_i$) of $\mathbb{Z}_{m_i}^*$ can be applied as an element-wise product over this isomorphic domain, and each different component represents a different irreducible subrepresentation of $V$.

\paragraph{Outer tensor product of Linear Representations}
Consider two groups $(G_1, \cdot)$ and $(G_2, \cdot)$ and consider the direct product $G_1 \times G_2$ with the following ``$\cdot$'' operation: $(s_1, s_2) \cdot (t_1, t_2) = (s_1 \cdot s_2, t_1 \cdot t_2)$ where $(s_1, s_2), (t_1, t_2) \in G_1 \times G_2$.

If we now define $\rho^{1}: G_1 \rightarrow \mbox{GL}(V_1)$ and $\rho^{2}: G_2 \rightarrow \mbox{GL}(V_2)$ as linear representations of $G_1$ and $G_2$, we can now define a linear representation $\rho^1 \otimes \rho^2: G_1 \times G_2 \rightarrow \mbox{GL}(V_1 \bigotimes V_2)$ by setting:
\begin{equation*}
\left( \rho^1 \otimes \rho^2 \right)(s_1, s_2) = \rho^1(s_1) \otimes \rho^2(s_2).
\end{equation*}

This way of dealing with the tensor of different linear representations allows us to define the different automorphisms of the tensor field $K_{(T)} = \bigotimes_{i \in \left[ l \right]} K_i$ in terms of the automorphisms of each $K_i$. Then, we have for $K_{(T)}$ the corresponding homomorphism with the tensor of linear representations $\bigotimes_{i \in \left[ l \right]} \rho_i: \bigoplus_{i \in \left[ l \right]} \mathbb{Z}_{m_i}^* \rightarrow \mbox{GL}\left( \bigotimes_{i \in \left[ l \right]} \mathbb{Q}(\varsigma_{m_i}) \right)$, and where each $\rho_i$ satisfies $\rho_i(k_i) = \uptau_{k_i}^{(i)}$, with $k_i \in \mathbb{Z}_{m_i}^*$ and being $\uptau_{k_i}^{(i)}$ the corresponding $\phi(m_i)$ automorphisms of the $K_i$ number field.

Finally, in order to map the set of $\prod_{i \in \left[ l \right]} \phi(m_i)$ automorphisms $\bigotimes_{i \in \left[ l \right]} \uptau_{k_i}^{(i)}$ with only one index we can consider the relation given in Equation \eqref{eq:indextensor} (Appendix~\ref{sec:Ht}), in such a way that $k_i \in \mathbb{Z}_{m_i}^* = g^{(i)}(\left[ \phi(m_i) \right])$ and $j_i = {(g^{(i)}(k_i))}^{-1}$. 

\subsection{Chinese Remainder Theorem}

In this section we explain why the CRT works over multivariate polynomial rings and how the use of the previously presented automorphisms affects the decomposition caused by the CRT.

First, consider $R = \mathcal{O}_{K_i} = \mathbb{Z}\left[ \varsigma_{m_i} \right]$, the ring of integers of a number field $\mathbb{Q}(\varsigma_{m_i})$ where $\varsigma_{m_i}$ is the $m_i$-th primitive root of unity. We know that if we work with the ideal $\langle q \rangle = qR$ and $q \in \mathbb{Z}$ is a prime, we have the following factorization $\langle q \rangle = \prod_{i} \mathfrak{q}_i^e$ where there are $\phi(m_i)/(ef)$ different $\mathfrak{q}_i$ of norm $q^f$ and we have $e = \phi(q')$ and $f$ is the minimum natural number that satisfies $q^f \equiv 1 \bmod{m_i/q'}$ with $q'$ the largest power of $q$ that divides $m_i$.

For each ideal, we have $\mathfrak{q}_j = \langle q, F_j(\varsigma_{m_i}) \rangle$ with $\Phi_{m_i}(x) = \prod_{j} {\left( F_j(x) \right)}^e$ being the factorization of $\Phi_{m_i}(x)$ modulo $q$. As explained in \cite{LPR13}, when we consider that $q \equiv 1 \bmod{m_i}$, both $e$ and $f$ are equal to $1$ and as we have an $m_i$-th primitive root of unity $w_i$ in $\mathbb{Z}_q$ we see that $\Phi_{m_i}(x) = \prod_{j \in \mathbb{Z}_{m_i}^*} (x - w_i^j)$. Therefore, we finally have $\langle q \rangle = \prod_{j \in \mathbb{Z}_{m_i}^*} \mathfrak{q}_j$ with $\mathfrak{q}_j = \langle q, \varsigma_{m_i} - w_i^j \rangle$. In addition, we know that we can use the automorphism $\uptau_{k}^{(i)}$ to exchange the contents between two different prime ideals $\mathfrak{q}_j$ of $qR$, that is, we can do $\uptau_{k}^{(i)}(\mathfrak{q}_j) = \mathfrak{q}_{j/k}$ (see Lemma $2.16$ in \cite{LPR13}).

Now, resorting to Lemma \ref{lem:CRT} in Appendix~\ref{sec:CRT}, we have an isomorphism from $\mathbb{Z}[\varsigma_{m_i}]/\langle q \rangle$ to $\bigoplus_{j \in \mathbb{Z}_{m_i}^*} \mathbb{Z}[\varsigma_{m_i}]/\langle q, \varsigma_{m_i} - w_i^j \rangle$, that is in fact also isomorphic to $\mathbb{Z}_q^{\phi(m_i)}$.

\paragraph{Multivariate extension}
We can see the multivariate case $R = \bigotimes_{i \in \left[ l \right]} \mathcal{O}_{K_i}$ as the tensor product between the previously considered univariate rings, that is, we have $\bigotimes_{i \in \left[ l \right]} \mathbb{Z}[\varsigma_{m_i}]/\langle q \rangle$ where $q$ has to satisfy $q \equiv 1 \bmod{m_i}$ for all $i \in \left[ l \right]$. Now, we know that it is isomorphic to the tensor product of the respective direct sum in terms of the different prime ideals $\bigotimes_{i \in \left[ l \right] } \left( \bigoplus_{j \in \mathbb{Z}_{m_i}^*} \mathbb{Z}[\varsigma_{m_i}]/\langle q, \varsigma_{m_i} - w_i^j \rangle \right)$ where we know that the tensor and direct product commute, therefore having
\begin{center}
  $\bigoplus_{j \in \left[ \prod_{i \in \left[ l \right]} \phi(m_i) \right]} \left( \bigotimes_{k \in \left[ l \right] } \mathbb{Z}[\varsigma_{m_k}]/\langle q, \varsigma_{m_k} - w_k^{j_k} \rangle \right)$,
  \end{center}
where the mapping between the set $\{j_1, \ldots, j_l \}$ and $j$ is defined by Equation \eqref{eq:indextensor}. This ring is in fact isomorphic to $\mathbb{Z}_q^{\prod_{i \in \left[ l \right]} \phi(m_i)}$.

Resorting to the ring isomorphism $\varsigma_{m_i} \rightarrow x_i$ for $i \in \left[ l \right]$ we have the expression $\bigoplus_{i \in \left[ l \right], j_i \in \mathbb{Z}_{m_i}^*} \mathbb{Z}_q\left[ x_1, \ldots, x_l \right]/ \langle x_1 - w_1^{j_1}, \ldots, x_l - w_l^{j_l} \rangle$. Now, thanks to the mapping introduced in Equation \eqref{eq:indextensor}, we consider $\mathfrak{q}_j = \mathfrak{q}_{j_1, \ldots, j_l} = \langle x_1 - w_1^{j_1}, \ldots, x_l - w_l^{j_l} \rangle$ with $j \in \left[ \prod_{i \in \left[ l \right]} \phi(m_i) \right]$. First, it can be easily shown that each $\mathfrak{q}_j$ is an ideal and, as there is an isomorphism from $\mathbb{Z}_q[x_1, \ldots, x_l]/\mathfrak{q}_j$ to the finite field $\mathbb{Z}_q$, $\mathfrak{q}_j$ is a maximal ideal and also a prime ideal because every maximal ideal over a ring is also a prime ideal.

In order to show that all the $\mathfrak{q}_j$ are comaximal ideals we have the following \textit{reductio ad absurdum} argument: consider two different maximal ideals $\mathfrak{q}_j$ and $\mathfrak{q}_k$ with $k \neq j$; by definition, $\mathfrak{q}_k + \mathfrak{q}_j$ is also an ideal; we have three possible cases: a)  $\mathfrak{q}_k + \mathfrak{q}_j = \mathfrak{q}_k$, b) $\mathfrak{q}_k + \mathfrak{q}_j = \mathfrak{q}_j$ and c) there is another maximal ideal $\mathfrak{q}_k + \mathfrak{q}_j$. The first two cases are not true because $\mathfrak{q}_k$ and $\mathfrak{q}_j$ are different, and the third case is impossible because each ideal is maximal, hence having $\mathfrak{q}_k + \mathfrak{q}_j = \mathbb{Z}_q[x_1, \ldots, x_l]$, which is the definition of comaximal ideals.

Then, knowing that we have a set of comaximal ideals $\mathfrak{q}_j$ for $j \in \left[ \prod_{i \in \left[ l \right]} \phi(m_i) \right]$, we can use Lemma \ref{lem:CRT} in Appendix~\ref{sec:CRT} to show that there exists an isomorphism from $\mathbb{Z}_q[x_1, \ldots, x_l]/\langle \Phi_{m_1}(x_1), \ldots, \Phi_{m_l}(x_l) \rangle$ to $\bigoplus_{j \in \left[ \prod_{i \in \left[ l \right]} \phi(m_i) \right]} \left( \mathbb{Z}_q[x_1, \ldots, x_l]/\mathfrak{q}_j \right)$, that is, we can compute the corresponding CRT, and the rest of the properties discussed in Appendix~\ref{sec:CRT} also apply.

Now, we can present a similar result to Lemma $2.16$ in \cite{LPR13}, but adapted to our more general case:
\begin{lem}[Lyubashevsky \emph{et al.} \cite{LPR13} Lemma $2.16$] 
For any $\mathfrak{q}_j = \mathfrak{q}_{j_1, \ldots, j_l}$ and $\mathfrak{q}_{j'} = \mathfrak{q}_{j'_1, \ldots, j'_l}$ (by Equation~\eqref{eq:indextensor}), we have a linear representation or automorphism $\otimes_{i \in \left[ l \right]} \rho_i \left( k_1, \ldots, k_l \right) = \otimes_{i \in \left[ l \right]} \uptau_{k_i}^{(i)}$ where $k_i \in \mathbb{Z}_{m_i}^*$ satisfies $\otimes_{i \in \left[ l \right]} \uptau_{k_i}^{(i)} \left( \mathfrak{q}_j \right) = \mathfrak{q}_{j'}$.
\end{lem}

\section{multivariate Ring-LWE}
\label{sec:mrlwe}

We define the multivariate RLWE distribution as a generalization of the RLWE distribution where the involved polynomial rings can have several indeterminates. The $m$-RLWE distribution is parameterized by a tensor of number fields $K_{(T)} = \bigotimes_{i \in \left[ l \right]} K_i$ where each $K_i$ is a cyclotomic number field; not necessarily being all of them different. We also consider the ring $R$ as the tensor of the corresponding ring of integers $\mathcal{O}_{K_i}$, that is, $R = \bigotimes_{i \in \left[ l \right]} \mathcal{O}_{K_i}$ and an integer modulus $q \geq 2$. We denote $\mathcal{J}_q$ for $\mathcal{J}/q\mathcal{J}$ where $\mathcal{J}$ is a fractional ideal in $K_{(T)}$. Let $R^{\vee}$ be the dual fractional ideal of $R$ and $\mathbb{T} = K_{(T), \mathbb{R}}/R^{\vee}$.\footnote{$K_{(T), \mathbb{R}}$ is defined as $K_{(T)} \bigotimes_{\mathbb{Q}} \mathbb{R}$. For more details we refer the reader to Appendix \ref{sec:embandgeom}.}

\begin{defin}[Multivariate ring LWE distribution] For $s \in R_q^{\vee}$ and an error distribution $\psi$ over $K_{(T), \mathbb{R}}$, a sample from the $m$-RLWE distribution $A_{s, \psi}$ over $R_q \times \mathbb{T}$ is generated by $a \leftarrow R_q$ uniformly at random, $e \leftarrow \psi$, and outputting $(a, b = (a \cdot s)/q + e \bmod{R^{\vee}})$.
\end{defin}

\begin{defin}[Multivariate ring LWE, Search\label{def:search}] Let $\Psi$ be a family of distributions over $K_{(T), \mathbb{R}}$. $m\mbox{-RLWE}_{q, \Psi}$ denotes the search version of the $m$-RLWE problem. It is defined as follows: given access to arbitrarily many independent samples from $A_{s, \Psi}$ for some arbitrary $s \in R_q^{\vee}$ and $\psi \in \Psi$, find $s$.
\end{defin}

Next, we include the decision version of the $m$-RLWE problem:

\begin{defin}[Multivariate ring LWE, Average-Case Decision\label{def:av-case}] Let $\Upsilon$ be a distribution over a family of error distributions, each over $K_{(T), \mathbb{R}}$. The average-case decision version of the $m$-RLWE problem, denoted $m\mbox{-R-DLWE}_{q, \Upsilon}$, is to distinguish with nonnegligible advantage between arbitrarily many independent samples from $A_{s, \psi}$, for a random choice of $(s, \psi) \leftarrow U(R_q^{\vee}) \times \Upsilon$,\footnote{$U(R_q^{\vee})$ represents the uniform distribution over $R_q^{\vee}$} and the same number of uniformly random and independent samples from $R_q \times \mathbb{T}$.
\end{defin}

For an asymptotic treatment of the $m$-RLWE problems, we let $K_{(T)}$ come from an infinite sequence of tensors of number fields $\mathbb{K} = \{ K_{(T), n} \}$ of increasing dimension $n$ ($n$ is the number of basis elements that form the integral basis), and let $q$, $\Psi$, and $\Upsilon$ depend on $n$ as well.

\paragraph{Error distributions}
We include here two definitions about the error distributions to achieve the reductions for the search version of multivariate ring-LWE (Definition~\ref{def:definitionerror1}) and for the hardness result for the average-case decision problem (Definition~\ref{def:definitionerror2}). We refer the reader to Appendices~\ref{sec:latticeBackground} and \ref{sec:embandgeom} for further information about Gaussian distributions over a tensor field.
 
\begin{defin}[extension of Lyubashevsky \emph{et al.} \cite{LPR13}, Definition $3.4$]\label{def:definitionerror1} For a positive real $\alpha > 0$, the family $\Psi_{\leq \alpha}$ is the set of all elliptical Gaussian distributions $D_{\bm{r}}$ (over $K_{(T), \mathbb{R}}$) where each parameter $r_i \leq \alpha$ with $i \in \left[ n \right]$.
\end{defin}

\begin{defin}[extension of Lyubashevsky \emph{et al.} \cite{LPR13}, Definition $3.5$]\label{def:definitionerror2} Let $K_{(T)} = \bigotimes_{i \in \left[ l \right]} K_i$ where the $K_i$ are the $m_i$-th cyclotomic number field having degree $n_i = \phi(m_i)$. For a positive real $\alpha > 0$, a distribution sampled from $\Upsilon_{\alpha}$ is given by an elliptical Gaussian distribution $D_{\bm{r}}$ (over $K_{(T), \mathbb{R}}$) whose parameters are $r_{i, j} = r_{i, j + n_i/2}$ (see Appendix~\ref{sec:gaussianmeasure}) and each $r_j$ with $j \in \left[ n \right]$ satisfies $r_j^2 = \alpha^2(1 + \sqrt{n}x_j)$, where whenever we have $r_i$ and $r_j$ such that $i, j \in \left[ n \right]$, $i \neq j$, the corresponding $x_i$ and $x_j$ are chosen independently from the distribution $\Gamma(2, 1)$.
\end{defin}

Our \emph{main theorem} is obtained by combining the theorems from Sections \ref{sec:hardlwe} and \ref{sec:pseudorandommrlwe} (see Appendix~\ref{sec:idealLatticeProblems} for the definitions of lattice hardness problems; i.e., SVP and SIVP):

\begin{thmRLWE}[Extended version to $m$-RLWE of Lyubashevsky \emph{et al.} \cite{LPR13} Theorem $3.6$ \label{Th:theorem36}] Let $K_{(T)} = \bigotimes_{i \in \left[ l \right]} K_i$ be the tensor product of $l$ cyclotomic fields of dimension $n_i = \phi(m_i)$ each, and $R = \bigotimes_{i \in \left[ l \right]} \mathcal{O}_{K_i}$ the tensor of their corresponding ring of integers. Let $\alpha < \sqrt{\log{n}/n}$, and let $q = q(n) \geq 2$, $q \equiv 1 \mod{m_i}$, for all $i$, be a poly$(n)$-bounded prime such that $\alpha q \geq \omega(\sqrt{\log{n}})$, where $\omega(f(n))$ denotes a function that asymptotically grows faster than $f(n)$. Then, there is a polynomial-time quantum reduction from $\tilde{\mathcal{O}}(\sqrt{n} / \alpha)$-approximate SIVP (or SVP) on (tensor) ideal lattices in $K_{(T)}$ to $m\mbox{-R-DLWE}_{q, \Upsilon_{\alpha}}$. Alternatively, for any $l \geq 1$, we can replace the target problem by the problem of solving $m\mbox{-R-DLWE}_{q, D_{\xi}}$ given only $l$ samples, where $\xi = \alpha \cdot {(nl/\log{nl})}^{1/4}$.
\end{thmRLWE}

\paragraph{Discretizing the $b$ component}

In practical applications \cite{PTP15}, we usually deal with a version of the hardness problem where the error distribution is discrete. That is, instead of working with an error distribution $\psi$ over $K_{(T), \mathbb{R}}$, we have to deal with an $m$-RLWE distribution $A_{s, \chi}$ where $\chi$ is a discrete error distribution over $R^{\vee}$ therefore resulting in an element $b$ that belongs to $R_{q}^{\vee}$.

Here, we present a variant of Definition \ref{def:av-case} that we call $m\mbox{-R-DLWE}_{q, \chi}$ where we have a given number of samples from $\chi$ instead of $\psi$, and we have the problem of distinguishing between samples from $A_{s, \chi}$ and uniform samples from $R_q \times R_q^{\vee}$.

In order to guarantee the hardness of the discrete version we have to follow the procedure described in \cite{LPR13b}. We include the main lemmas that explain the hardness of the discrete version together with some relevant explanations about the considerations needed for our multivariate case.

The following lemma states that if $m\mbox{-R-DLWE}_{q, \psi}$ is hard with $l$ samples, then $m\mbox{-R-DLWE}_{q, \chi}$ is also hard for the same number of samples, with $\chi$ the distribution obtained from ${\lfloor p \cdot \psi \rceil}_{w + pR^{\vee}}$ and $p$ and $q$ coprime integers.

\begin{lem}[Extended version of Lemma $2.23$ in \cite{LPR13b}] Let $p$ and $q$ coprime integers, and $\lfloor \cdot \rceil$ a valid discretization to cosets of $pR^{\vee}$. There exists an efficient transformation that on input $w \in R_p^{\vee}$ and a pair in $(a', b') \in R_q \times K_{(T), \mathbb{R}}/qR^{\vee}$ outputs $(a = pa' \bmod{qR}, b) \in R_q \times R_q^{\vee}$ with the following considerations: if the input pair is uniformly distributed then so is the output pair; and if the input pair is distributed according to the multivariate ring-LWE distribution $A_{s, \psi}$ for some unknown $s \in R^{\vee}$ and distribution $\psi$ over $K_{(T), \mathbb{R}}$, then the output is distributed according to $A_{s, \chi}$ where we have that $\chi = {\lfloor p \cdot \psi \rceil}_{w + pR^{\vee}}$.
\end{lem}

In practical applications \cite{PTP15} it is also common to have two additional changes with respect to the previous definition of the average-case decision version: a) instead of sampling $a$ and $s$ from $R_q$ and $R_q^{\vee}$ respectively, both are usually sampled from $R_q$. In general, we are in a different situation when we do this, however the works which consider that $s$ belongs to $R_q$ deal with a particular type of cyclotomic fields where $m_i$ is a power of two. It can be shown that for this particular type of cyclotomic fields both definitions are equivalent, so it does not introduce additional drawbacks to the hardness reduction; b) instead of a uniform $s$, $s$ is chosen from the error distribution (this is known as ``normal form'') in practical cases, hence having a short secret key.

In order to show that the variant with short error ($\mbox{R-DLWE}_{q, \chi}$) is as hard as the original $\mbox{R-DLWE}_{q, \psi}$, the proof of Lyubashevsky \emph{et al.} \cite{LPR13b} follows the technique of \cite{ACPS09}. Their results can be adapted to our more general case, so we include below the relevant lemma:

\begin{lem}[Extended version of Lemma $2.24$ in \cite{LPR13b}\label{lemma12}] Let $p$ and $q$ be positive coprime integers, $\lfloor \cdot \rceil$ be a valid discretization to cosets of $pR^{\vee}$, and $w$ be an arbitrary element in $R_p^{\vee}$. If $m\mbox{-R-DLWE}_{q, \psi}$ is hard given some number $l$ of samples, then so is the variant of $m\mbox{-R-DLWE}_{q, \chi}$ where the secret is sampled from $\chi = {\lfloor p \cdot \psi \rceil}_{w + pR^{\vee}}$, given $l - 1$ samples.
\end{lem}

The proof of the previous lemma relies on how to use an oracle of the second problem to solve the first one. The difference with respect the proof presented in \cite{LPR13b} lies on how to compute the fraction of invertible elements of $R_q$. In order to resolve this, we resort to the following claim about cyclotomic fields:
\begin{claim}[Claim $2.25$ in \cite{LPR13b}] Consider the $m$-th cyclotomic field of degree $n = \phi(m)$ for some $m \geq 2$. Then for any $q \geq 2$, the fraction of invertible elements in $R_q$ is at least $1/\mbox{poly}\left( n, \log{q} \right)$.
\end{claim}

In our case, we work with the tensor of cyclotomic fields $K_{(T)} = \bigotimes_{i \in \left[ l \right]} K_i$; for each cyclotomic field $K_i$, the fraction of irreducible elements in $\mathcal{O}_{K_i}/\langle q \rangle$ is at least $1/\mbox{poly}\left( \phi(m_i), \log{q} \right)$ with $q \geq 2$ and with $q \equiv 1 \bmod{m_i}$ for all $i \in \left[ l \right]$. 
When working in the tensor of the different polynomial rings over $\mathbb{Z}_q$, if an element is invertible, the corresponding elements belonging to each $\mathcal{O}_{K_i}$ must be invertible too (same explanation as in Kronecker product of matrices, Section~\ref{specialembedding}). Then, the fraction of invertible elements in $R_q = \bigotimes_{i \in \left[ l \right]} \mathcal{O}_{K_i}/\langle q \rangle$ is at least the product of the fractions of each ring of integers $1/\mbox{poly}\left( \prod_{i \in \left[ l \right]} \phi(m_i), \log{q} \right) = 1/\mbox{poly}\left( n, \log{q} \right)$, and Lemma \ref{lemma12} follows.

\section{Proof sketch of the hardness of the multivariate Ring Learning with Errors problem}
\label{sec:proofsketch}

This section introduces the main theorems together with their proofs for the different reductions of the $m$-RLWE problem. The proof can be divided in two main parts, described in the following paragraphs.

\paragraph*{Hardness Search-LWE} The first part achieves a quantum reduction from approximate SVP on ideal lattices over $R$ to the search version of $m$-RLWE. The goal of the search version is to recover the secret key $s$. The procedure follows the techniques considered by Lyubashevsky \emph{et al.} \cite{LPR13} and Regev \cite{Regev09}.

The main contribution here is to extend their tools to the more general case of the tensor of cyclotomic fields (or even the tensor of more general fields). For this purpose, we use the iterative quantum reduction for general lattices of Regev together with the corresponding tools that we can find on algebraic number theory; i.e., the Chinese Remainder Theorem and the canonical embedding that were used by Lyubashevsky \emph{et al.} but adapted to our multivariate case.

\paragraph*{Pseudorandomness of $m$-RLWE} The main purpose of this part is to show that the $m$-RLWE distribution is pseudorandom, that is, there exists a reduction from the search problem, discussed in the first part, to the decision variant of the hardness problem. We present two different versions of the hardness problem: one for the decision problem with a nonspherical distribution in the canonical embedding, and another one for the decision problem with a spherical distribution but with a bounded number of samples. Additionally, when assuming the hardness of the search problem with a fixed spherical Gaussian error distribution, we also have hardness of the decision version with the same error distribution.

Again, the main contribution of our work relies on proving that the multivariate samples following the $m$-RLWE distribution are pseudorandom, therefore generalizing the results of \cite{LPR13} to the case of multivariate elements.
The main needed properties are those related to the decomposition of $\langle q \rangle$ into $n$ prime ideals and the use of the automorphisms allowing us to permute the prime ideals.

\subsection{Hardness Search-LWE}
\label{sec:hardlwe}

For this section, let $K_{(T)} = \bigotimes_{i \in \left[ l \right]} K_i$ of degree $n$ denote the tensor of $l$ arbitrary number fields and $R = \bigotimes_{i \in \left[ l \right]} \mathcal{O}_{K_i}$ the corresponding tensor of rings of integers. The results can be applied to an arbitrary number field, so in this section we do not have to consider the specific case of cyclotomic fields.

\begin{thmRLWE}[Extended Theorem $4.1$ of Lyubashevsky \emph{et al.} \cite{LPR13}\label{th:hardsearchLWE}] Let $K_{(T)}$ be a tensor of arbitrary number fields with degree $n_i$ each and $R$ the tensor of the corresponding ring of integers. Let $\alpha = \alpha(n) > 0$, and let $q = q(n) \geq 2$ be such that $\alpha q \geq 2 \cdot \omega(\sqrt{\log{n}})$, where $\omega(f(n))$ denotes a function that asymptotically grows faster than $f(n)$. For some negligible $\epsilon = \epsilon(n)$, there is a probabilistic polynomial-time quantum reduction from $\mbox{K}_{(T)}\mbox{-DGS}_{\gamma}$ to $m\mbox{-R-LWE}_{q, \Psi_{\leq \alpha}}$, where
  \begin{center}
    $\gamma = \max{ \{ \eta_{\epsilon}(\mathcal{I}) \cdot (\sqrt{2} / \alpha) \cdot \omega(\sqrt{\log{n}}), \sqrt{2n}/\lambda_{1}(\mathcal{I}^{\vee}) \}}$
  \end{center}
\end{thmRLWE}

Here $\mbox{K}_{(T)}\mbox{-DGS}_{\gamma}$ denotes the discrete Gaussian sampling problem \cite{Regev09,LPR13} where given an ideal $\mathcal{I}$ in $K_{(T)}$ and a number $s \geq \gamma = \gamma(\mathcal{I})$, we have to generate samples from $D_{\mathcal{I}, s}$. The proof of this theorem is shown in Appendix~\ref{sec:proofhardsearchLWE}.

Regev~\cite{Regev09} showed that we have easy reductions from standard lattice problems to $\mbox{DGS}$. As Lyubashevsky \emph{et al.}~\cite{LPR13} assert, combining lemmas \ref{lem:2.2} and \ref{lem:2.4} we have $\eta_{\epsilon}(\mathcal{I}) \leq \lambda_n(\mathcal{I}) \cdot \omega(\sqrt{\log{n}})$ (see Appendix \ref{sec:latticeBackground} for the definition of the smoothing parameter $\eta_{\epsilon}$) for any fractional ideal $\mathcal{I}$ and negligible $\epsilon(n)$, and we also have that samples from $D_{\mathcal{I}, \gamma}$ have length at most $\gamma \sqrt{n}$ with overwhelming probability. This is also valid in our case.

Analogously, an oracle for $\mbox{K}_{(T)}\mbox{-DGS}_{\gamma}$ with $\gamma = \eta_{\epsilon}(\mathcal{I}) \cdot \tilde{\mathcal{O}}(1 / \alpha)$ implies an oracle for $\tilde{\mathcal{O}}(\sqrt{n}/\alpha)$-approximate SIVP on ideal lattices in the tensor field $K_{(T)}$.

When each $K_i$ is a cyclotomic field, we also have $\lambda_n(\mathcal{I}) = \lambda_1(\mathcal{I})$ for any fractional ideal $\mathcal{I}$, as for each shortest nonzero $v \in \mathcal{I}$, if we multiply it by different combinations of $\varsigma_{m_1}^{e_1 - 1} \otimes \ldots \otimes \varsigma_{m_l}^{e_l - 1}$ with $e_i \in \left[ \phi(m_i) \right]$, it yields a total of $n$ independent elements of equal length, that is, we have an oracle for $\tilde{\mathcal{O}}(\sqrt{n}/\alpha)$-approximate SVP.

It is important to note that as the error distribution is added modulo $R^{\vee}$ in the definition of $m$-RLWE, the condition $\alpha < \eta_{\epsilon}(R^{\vee})$ must be satisfied for all negligible $\epsilon(n)$ for the problem to be solvable.

\subsection{Pseudorandomness of $m$-RLWE}
\label{sec:pseudorandommrlwe}

In this section, we particularize again $K_{(T)} = \bigotimes_{i \in \left[ l \right]} K_i$ and $R = \bigotimes_{i \in \left[ l \right]} \mathcal{O}_{K_i}$ for the cyclotomic case $K_i = \mathbb{Q}(\varsigma_{m_i})$ with $\varsigma_{m_i}$ a primitive $m_i$-th root of unity. 
We also consider the prime $q \equiv 1 \bmod{m_i}$ for all $i \in \left[ l \right]$ and we have that it is $\mbox{poly}(n)$-bounded, where $n = \prod_{i \in \left[ l \right]} \phi(m_i)$ is the degree of the considered multivariate polynomials.

We recall that $K_{(T)}$ has a set of $n$ different automorphisms $\uptau_{j}$ with $j \in \left[ n \right]$ (see Equation \eqref{eq:indextensor}) and when working over $q$, we have that $\langle q \rangle = \prod_{i \in \left[ n \right]} \mathfrak{q}_i$ splits into a product of prime ideals $\mathfrak{q}_i$ where the automorphisms satisfy $\otimes_{i \in \left[ l \right]} \uptau_{k_i}^{(i)}\left( \mathfrak{q}_j \right) = \mathfrak{q}_{j'}$ for any prime ideals $\mathfrak{q}_j$, $\mathfrak{q}_{j'}$ where $k_i \in \mathbb{Z}_{m_i}^*$ and $j,j' \in \left[ n \right]$ (for more details we refer the reader to Appendix~\ref{sec:prelim}). 

In the following we present the main theorems about the different reductions from the search version of $m$-RLWE (see Definition~\ref{def:search} and Theorem~\ref{th:hardsearchLWE} about the reduction over worst-case lattice problems) to the average-case decision problem $m\mbox{-R-DLWE}$ (see Definition \ref{def:av-case}).

\begin{thmRLWE}[Extended Theorem $5.1$ of Lyubashevsky \emph{et al.} \cite{LPR13}\label{th:pseudorandomness1}] Let $R$ and $q$ be as shown previously and let $\alpha q \geq \eta_{\epsilon}\left( R^{\vee} \right)$ for some negligible $\epsilon = \epsilon(n)$. Then, there is a randomized polynomial-time reduction from $m\mbox{-R-LWE}_{q, \Psi_{\leq \alpha}}$ to $m\mbox{-R-DLWE}_{q, \Upsilon_{\alpha}}$.
\end{thmRLWE}

In order to prove the previous theorem we need four more reductions that are described in the following discussion.
\begin{align*} \label{diagram1}
  \mbox{LWE}_{q, \Psi} \xrightarrow[\mbox{Lemma \ref{lem:5.5}}]{Automorphisms} \mathfrak{q}_i\mbox{-LWE}_{q, \Psi} \xrightarrow[\mbox{Lemma \ref{lem:5.9}}]{Search/Decision} \mbox{WDLWE}_{q, \Psi}^{i}\\
\mbox{WDLWE}_{q, \Psi}^{i}\xrightarrow[\mbox{Lemma \ref{lem:5.12}}]{Worst/Average} \mbox{DLWE}_{q, \Upsilon}^{i} \xrightarrow[\mbox{Lemma \ref{lem:5.14}}]{Hybrid} \mbox{DLWE}_{q, \Upsilon}
\end{align*} 

The details of the proof follow the steps of Lyubashevsky \emph{et al.} \cite{LPR13}, which, conversely, follows similar steps to the reductions of \cite{Regev09}, the main point being the use of the automorphisms to recover the secret key $s$ when only knowing the secret key relative to one prime ideal $\mathfrak{q}_i$ (Lemma \ref{lem:5.5}).

An additional needed step is the randomization of the error distribution (sampled from $\Upsilon$) such that the error is invariant under the different field automorphisms (see Lemma \ref{lem:5.12}) because the different $\psi \in \Psi_{\leq \alpha}$ are not necessarily invariant under the field automorphisms. Equivalently, if this reduction randomizing the error distribution is not desirable, we can apply a bound on the number of samples for considering a result about pseudorandomness of $m$-RLWE with a fixed spherical noise distribution.

\begin{thmRLWE}[Extended Theorem $5.2$ of Lyubashevsky \emph{et al.} \cite{LPR13}\label{th:pseudorandomness2}] Let $R$, $q$ and $\alpha$ be as in Theorem \ref{th:pseudorandomness1} and let $l \geq 1$. There is a randomized polynomial-time reduction from solving $m\mbox{-R-LWE}_{q, \Psi_{\leq \alpha}}$ to solving $m\mbox{-R-DLWE}_{q, D_{\xi}}$ given only $l$ samples, where $\xi = \alpha \cdot {\left( nl/\log{(nl)} \right)}^{1/4}$.
\end{thmRLWE}
In this case, we have a similar reduction to the one in Theorem \ref{th:pseudorandomness1} but considering a different lemma (Lemma \ref{lem:5.16} instead of Lemma \ref{lem:5.12} in one of the steps).
\begin{equation*}\label{diagram2}
  \mbox{WDLWE}_{q, \Psi}^{i}\xrightarrow[\mbox{Lemma \ref{lem:5.16}}]{Worst/Average} \mbox{DLWE}_{q, D_{\xi}}^{i} \xrightarrow[\mbox{Lemma \ref{lem:5.14}}]{Hybrid} \mbox{DLWE}_{q, D_{\xi}}
\end{equation*} 

It is interesting to note that if we assume hardness of the search version with a spherical error distribution $\mbox{LWE}_{q, D_{\xi}}$, then we also have a reduction for the pseudorandomness with a spherical error, but simplifying Lemma \ref{lem:5.12} instead of resorting to sampling from the $\Upsilon$ distribution.

\begin{thmRLWE}[Extended Theorem $5.3$ of Lyubashevsky \emph{et al.} \cite{LPR13}\label{th:pseudorandomness3}] Let $R$, $q$ and $\alpha$ be as in Theorem \ref{th:pseudorandomness1}. There exists a randomized polynomial-time reduction from solving $m\mbox{-R-LWE}_{q, D_{\alpha}}$ to solving $m\mbox{-R-DLWE}_{q,D_{\alpha}}$.
\end{thmRLWE}

The detailed proofs for these three theorems along with the lemmas involved in the security reductions for $m$-RLWE are included in Appendix~\ref{sec:proofs4everyting}.

\section{Conclusions}
In this work we have presented a multivariate version of the well-known Ring Learning with Errors (RLWE) problem to a multivariate version working over the tensor product of number fields, denoted $m$-RLWE, which finds application in secure signal processing scenarios. We have adapted and generalized the techniques of Lyubashevsky~\emph{et al.}~\cite{LPR13} to the tensor product of number fields and achieved a reduction of the $m$-RLWE problem to hardness problems over ideal lattices, hence giving some new insights into its security.

\bibliographystyle{splncs03}
\bibliography{biblio}

\appendix

\section{Fundamental Concepts of Lattices and Algebraic Number Theory}
\label{sec:prelim}

This appendix presents the fundamental concepts of lattices and algebraic number theory and extends them to the more general case of a tensor of number fields on which $m$-RLWE is mainly based.

\subsection{The Space $H_{(T)} = \bigotimes_i H_i$}
\label{sec:Ht}

When working with cyclotomic fields, it is useful to work with the subspace $H \subseteq \mathbb{R}^{s_1} \times \mathbb{C}^{2s_2}$ with $s_1 + 2s_2 = n$, where the tuple $(s_1, s_2)$ is called the signature of the number field, and $H$ satisfies:

\begin{equation}
H = \{ (x_1, \ldots, x_n) \in \mathbb{R}^{s_1} \times \mathbb{C}^{2s_2} \mbox{ such that } x_{s_1 + s_2 + j} = \bar{x}_{s_1 + j}, \forall j \in [ s_2 ] \} \subseteq \mathbb{C}^n
\end{equation}

An orthonormal basis $\{ \bm{h}_j \}_{j \in [ n ]}$ for $H$ can be defined as:

\begin{equation}
\bm{h}_j =
\left\{
	\begin{array}{cc}
	   \bm{e}_j & \mbox{if } j \in [ s_1 ] \\
	   \frac{1}{\sqrt{2}} (\bm{e}_j + \bm{e}_{j + s_2}) & \mbox{if } s_1 < j \leq s_1 + s_2 \\
           \frac{\sqrt{-1}}{\sqrt{2}} (\bm{e}_{j - s_2} - \bm{e}_{j}) & \mbox{if } s_1 + s_2 < j \leq s_1 + 2s_2
	\end{array}
\right.
\end{equation}
where the vectors $\bm{e}_j$ are the vectors of the standard basis in $\mathbb{R}^n$.

Finally, each element $a = \sum_{j \in \left[ n \right]} a_j\bm{h}_j \in H$ (where all $a_j \in \mathbb{R}$) has its own $l_p$ norm defined as in Section~\ref{sec:not}.

For our purposes, we define the subspace $H_{(T)} = \bigotimes_{i \in \left[ l \right]} H_i$ as the tensor product of $l$ subspaces $H_i$, each equivalent to the subspaces previously introduced.

In particular, if we see each element belonging to each $H_i$ as a different linear transformation, we are actually working with the Kronecker product of the different subspaces $H_i$. Hence, the new basis will be the result of the Kronecker product of the original basis of each $H_i$, therefore having an orthonormal basis for $H_{(T)}$ given by $\{ \bm{h}_j \}_{j \in \left[ n \right]}$, where we can define the following mapping for $j$
\begin{equation}
\label{eq:indextensor}
j = 1 + \sum_{i \in \left[ l \right] } \left( j_i - 1 \right) \prod_{d \in \left[ i \right]} n_{d - 1},
\end{equation}
being $\bm{h}_j = \bigotimes_{i \in \left[ l \right]} \bm{h}_{j_i}^{(i)}$ the new form of the basis vectors, and where $n = \prod_{i \in \left[ l \right]} n_i$ and each $\{ \bm{h}_{j_i}^{(i)}\}_{j_i \in \left[ n_i \right]}$ is the corresponding orthonormal basis of each $H_i \subseteq \mathbb{C}^{n_i}$ for $i \in \left[ l \right]$ and $n_0 = 1$. This expression is used when indexing the embeddings (see Appendix~\ref{sec:embandgeom}) and automorphisms (see Section \ref{specialproperties}) that can be performed in a tensor field.

\subsection{Lattice background}
\label{sec:latticeBackground}

A lattice in our multivariate setting is defined as an additive subgroup of $H_{(T)} = \bigotimes_{i \in \left[ l \right]} H_i$. We only work with lattices of full rank, which are obtained as the set of all integer linear combinations of a set of $n$ linear independent basis vectors $\bm{B} = \{ \bm{b}_1, \ldots, \bm{b}_n \} \subset H_{(T)}$:\footnote{As we work with the Kronecker product of a basis for each subspace $H_i$, we can exploit the properties of the Kronecker product to find the corresponding basis.}

\begin{equation}
\Lambda = \mathcal{L}(B) = \left\{ \sum_{i \in [ n ]} z_i \bm{b}_i \mbox{ such that } \bm{z} \in \mathbb{Z}^n \right\}
\end{equation}

The minimum distance $\lambda_1(\Lambda)$ of a lattice $\Lambda$ for the norm $||.||$ is given with the length of the shortest nonzero lattice vector, that is, $\lambda_1(\Lambda) = \mbox{min}_{\bm{x} \in \Lambda / \bm{x} \neq \bm{0}} || \bm{x} ||$.

The dual lattice of $\Lambda \subset H_{(T)}$ is defined as $\Lambda^* = \{ \bm{x} \in H_{(T)} \mbox{ such that } \langle \Lambda, \bm{x} \rangle \subseteq \mathbb{Z} \}$ and it satisfies ${(\Lambda^*)}^* = \Lambda$.

\paragraph{Gaussian Measures}
\label{sec:gaussianmeasure}

The results explained in \cite{LPR13} for nonspherical Gaussian distributions can be extended to our case. So we repeat here some of the concepts presented for Gaussian measures but adapted to our tensor setting.

We consider the Gaussian function $\rho_r: H \rightarrow (0, 1]$ with $r > 0$ as $\rho_r(\bm{x}) = \mbox{exp}(-\pi {|| \bm{x} ||}^2 /r^2)$. A continuous Gaussian probability distribution can be obtained by normalizing the previous function in such a way that we have $D_r$ with a density function $r^{-n}\rho_r(\bm{x})$. When we extend this to the non spherical Gaussian case, we consider the vector $\bm{r} = \bigotimes_{i \in \left[ l \right]} \bm{r}_i$ where $\bm{r} = (r_1, \ldots, r_n)\in {(\mathbb{R}^+)}^{n}$ or also each $\bm{r}_i = (r_{i,1}, \ldots, r_{i,n_i}) \in {(\mathbb{R}^+)}^{n_i}$ and whose components satisfy $r_{i,j + s_1 + s_2} = r_{i,j + s_1}$. Finally, a sample from $D_{\bm{r}}$ is given by $\sum_{i \in [ n ]} x_i \bm{h}_i$ where $x_{j} = \prod_{i \in \left[ l \right]} x_{j_i}^{(i)}$ and each $x_j$ is drawn independently from the Gaussian distribution $D_{r_j}$ over $\mathbb{R}$; being $r_j$ equal to $\prod_{i \in \left[ l \right]} r_{i,j_i}$ and using the mapping between $\{ j \}_{j \in \left[ n \right]}$ and $\{ j_i \}_{j_i \in \left[ n_i \right], i \in \left[ l \right]}$ given by equation \eqref{eq:indextensor}.

Next, we include several results about the Gaussian distributions that are needed for this work.

\begin{defin}[Smoothing parameter] The smoothing parameter $\eta_{\epsilon}(\Lambda)$ for a lattice $\Lambda$ and real $\epsilon > 0$ is defined as the smallest $r$ such that $\rho_{1/r}(\Lambda^{*} \backslash \{ \bm{0} \}) \leq \epsilon$.
\end{defin}

In addition, several important lemmas from \cite{LPR13}, \cite{MR07}, \cite{Regev09} and \cite{Banaszczyk93} about the relation between the smoothing parameter and properties of lattices are included below.

\begin{lem}[Lyubashevsky \emph{et al.} \cite{LPR13} Lemma $2.2$, Micciancio and Regev \cite{MR07} Lemmas $3.2$ and $3.3$\label{lem:2.2}] For any $n$-dimensional lattice $\Lambda$, we have $\eta_{2^{-2n}}(\Lambda) \leq \sqrt{n}/\lambda_1(\Lambda^*)$ and $\eta_{\epsilon}(\Lambda) \leq \sqrt{\ln(n/\epsilon)} \lambda_n(\Lambda)$ for all $0 < \epsilon < 1$.
\end{lem}

\begin{lem}[Lyubashevsky \emph{et al.} \cite{LPR13} Lemma $2.3$, Micciancio and Regev \cite{MR07} Lemma $4.1$, Regev \cite{Regev09} Claim $3.8$\label{lem:2.3}] For any lattice $\Lambda$, $\epsilon > 0$, $r \geq \eta_{\epsilon}(\Lambda)$, and $\bm{c} \in H_{(T)}$, the statistical distance\footnote{The statistical distance $\Delta(X, Y)$ between two continuous random variables $X$ and $Y$ over $\mathbb{R}^n$ with probability density functions $T_1$ and $T_2$ is defined as $\Delta(X, Y) = \frac{1}{2}\int_{\mathbb{R}^n} |T_1(r) - T_2(r)|dr$. For more details we refer the reader to \cite{MR07} and \cite{Regev09}.} between $(D_r + \bm{c}) \mod \Lambda$ and the uniform distribution modulo $\Lambda$ is at most $\epsilon/2$. Alternatively, we have $\rho_r(\Lambda + \bm{c}) \in \left[ \frac{1 - \epsilon}{1 + \epsilon}, 1 \right] \rho_r(\Lambda)$.
\end{lem}

Let a lattice $\Lambda$, a point $\bm{u} \in H_{(T)}$ and $r > 0$ with $r \in \mathbb{R}$, the discrete Gaussian probability distribution over $\Lambda + \bm{u}$ with parameter $r$ can be defined as $D_{\Lambda + \bm{u}, r}(\bm{x}) = \frac{\rho_r(\bm{x})}{\rho_r(\Lambda + \bm{u})}$ for all $\bm{x} \in \Lambda + \bm{u}$.

\begin{lem}[Banaszczyk \cite{Banaszczyk93}, Lemma $1.5$ (i)\label{lem:2.4}] For any $n$-dimensional lattice $\Lambda$ and $r > 0$, a sample point from $D_{\Lambda, r}$ has Euclidean norm at most $r\sqrt{n}$, except with probability at most $2^{-2n}$.
\end{lem}

\begin{lem}[Regev \cite{Regev09}]: Let $\Lambda$ be a lattice, let $\bm{u} \in H_{(T)}$ be any vector, and let $r, s > 0$ be reals. Assume that $1/\sqrt{1/r^2 + 1/s^2} \geq \eta_{\epsilon}(\Lambda)$ for some $\epsilon < 1/2$. Consider the continuous distribution $Y$ on $H_{(T)}$ obtained by sampling from $D_{\Lambda + \bm{u}, r}$ and then adding an element drawn independently from $D_s$. Then, the statistical distance between $Y$ and $D_{\sqrt{r^2 + s^2}}$ is at most $4\epsilon$.
\end{lem}

\subsection{Algebraic Number Theory background}
\label{sec:algnumbertheory}

This appendix covers the main concepts related to number fields that are used in the papers \cite{LPR13} and \cite{LPR13b}; we highlight the theorems and lemmas that are fundamental to our proof, so even when they have already been presented in the literature, we include them here for completeness and to make our work self-contained. We also particularize some of the results to the case of cyclotomic fields; for further details, we refer the reader to the previous cited papers or to any introductory book on the subject \cite{LiNi94}.

The concepts about algebraic number theory presented here are necessary to show which are the main changes needed to extend the proof of Lyubashevsky \emph{et al.} to the generic multidimensional case (not only coprime factors), as explained in Section~\ref{sec:proofsketch}.

\subsubsection{Number fields}

A number field is defined as a field extension $K = \mathbb{Q}(\varsigma)$ where the element $\varsigma$ is incorporated to the field of rationals. This element $\varsigma$ satisfies $f(\varsigma) = 0$ for an irreducible polynomial $f(x) \in \mathbb{Q}[x]$ denoted minimal polynomial of $\varsigma$. The degree $n$ of a number field is the degree of its minimal polynomial.

We can also see the number field $K$ as an $n$-dimensional vector space over $\mathbb{Q}$ where $\{ 1, \varsigma, \ldots, \varsigma^{n - 1} \}$ is called the power basis of the field $K$. Of course, we have an isomorphism between $K$ and $\mathbb{Q} \left[ x \right] / f(x)$.

In this work, we have a special interest on cyclotomic fields, which are those fields where $\varsigma = \varsigma_m$ (for some natural number $m$) is an $m$-th primitive root of unity and the minimal polynomial of $\varsigma_m$ is the $m$-th cyclotomic polynomial $\Phi_m(x) = \prod_{i \in \mathbb{Z}_m^*} (x - \omega_m^i) \in \mathbb{Z} \left[ x \right]$, where $\omega_m \in \mathbb{C}$ is any primitive $m$-th complex root of unity (for example $\omega_m = e^{2 \pi \sqrt{-1}/m}$). It is important to note that the different powers $\omega_m^i$ of $\Phi_m(x)$ are the $m$-th roots of unity in $\mathbb{C}$ and that the degree of $\Phi_m(x)$ is $n = \phi(m)$, where $\phi(m)$ is the Euler's totient function.

In general, there is no bound on the number of elements that can be added, so we could have $K = \mathbb{Q}(\varsigma_{m_1}, \ldots, \varsigma_{m_l})$, that is isomorphic to the cyclotomic field $\mathbb{Q}(\varsigma_m) = \bigotimes_{i \in \left[ l \right]} \mathbb{Q}(\varsigma_{m_i})$ when $m = \prod_{i \in \left[ l \right]} m_i$ has a prime-power decomposition and each $\varsigma_{m_i}$ is a $m_i$-th primitive root of unity (See \cite{LPR13b}).

Therefore, we can see our scheme as a generalization of the previous tensor product of cyclotomic fields, where we can have a non prime tensor decomposition of $m$ (the same power cyclotomic can appear several times in the expression).

\subsubsection{Embeddings and Geometry}
\label{sec:embandgeom}

Here, we describe the embeddings that can be defined in a general number field together with the canonical geometry that we can consider thanks to these embeddings.

A number field $K = Q(\varsigma)$ of degree $n$ has exactly $n$ embeddings $\sigma_i: K \rightarrow \mathbb{C}$ where each of these embeddings maps $\varsigma$ to a different complex root of its minimal polynomial $f$. The number of real embeddings is denoted $s_1$ and the number of pairs of complex embeddings is denoted by $s_2$ (each complex root has a conjugate), so we have $n = s_1 + 2s_2$ (the pair $(s_1, s_2)$ is called the signature of the number field).

The canonical embedding is defined as $\sigma: K \rightarrow \mathbb{R}^{s_1} \times \mathbb{C}^{2s_2}$ where $\sigma(x) = {\left( \sigma_1(x), \ldots, \sigma_n(x) \right)}^{T}$. We let $\{ \sigma_i \}$ with $i = 1, \ldots, s_1$ be the real embeddings and $\sigma_{s_1 + s_2 + j} = \bar{\sigma}_{s_1 + j}$ with $j = 0, \ldots, s_2 - 1$ be the complex embeddings.

For our purposes it is useful to redefine the embedding of $\bigotimes_{i \in \left[ l \right]} K_i$ as in \cite{LPR13b} with the corresponding reordering of the $\sigma_i(x)$. Therefore, we have $\sigma(\otimes_{i \in \left[ l \right]} a_i) = \otimes_{i \in \left[ l \right]} \sigma^{(i)}(a_i)$ and instead of considering the signature $\left( s_1, s_2 \right)$, each $\sigma^{(i)}$ is defined as $\sigma^{(i)}: K_i \rightarrow \mathbb{C}^{\mathbb{Z}_{m_i}^*}$ (for the particular case of cyclotomic fields with $m_i > 2$ there are no real roots, so we have $s_1 = 0$).

Now, we have a bijective map $g^{(i)} : \left[ \phi(m_i) \right] \rightarrow \mathbb{Z}_{m_i}^*$ that allows us to represent each embedding with a new set of indices as
\begin{equation*}\sigma^{(i)}(x) = {\left(\sigma_{g^{(i)}(1)}^{(i)}(x), \ldots, \sigma_{g^{(i)}(\phi(m_i))}^{(i)}(x) \right)}^{T}
\end{equation*}
in such a way that if $k_i \in \mathbb{Z}_{m_i}^* = g^{(i)}(\left[ \phi(m_i) \right])$, the relation between the complex conjugates is $\sigma^{(i)}_{k_i} = \bar{\sigma}^{(i)}_{m_i - k_i}$. Finally, the tensoring of the different embeddings $\otimes_{i \in \left[ l \right]} \sigma^{(i)}(a_i)$ reduces over $H_{(T)}$ in a Kronecker product of the images obtained in each different subspace $H_i$.

By virtue of this canonical embedding, there exists a ring homomorphism from $\bigotimes_{i \in \left[ l \right]} K_i$ to $\bigotimes_{i \in \left[ l \right]} H_i$ where each $H_i \subset \mathbb{C}^{\mathbb{Z}_{m_i}^*}$, and where multiplication and addition are element-wise. Thanks to this, we can define geometric norms over $\bigotimes_{i \in \left[ l \right]} K_i$ considering the presented tensor subspace $H_{(T)}$. Therefore, for any $x \in K_{(T)}$ and any $p \in [1, \infty ]$, we consider ${||x||}_p = {||\sigma(x)||}_p = { \left( \sum_{j \in [ n ] }{|\sigma_j(x)|}^p \right) }^{1/p}$ with $p < \infty$ and $\mbox{max}_{j \in [ n ]} |\sigma_j(x)|$ for $p = \infty$, where each $\sigma_j(x) = \prod_{i \in \left[ l \right]} \sigma_{g^{(i)}(j_i)}^{(i)}(x)$ following the mapping indicated in Equation \eqref{eq:indextensor}, and $j_i \in \left[ \phi(m_i) \right]$, $j \in \left[ n \right]$ such that $n = \prod_{i \in \left[ l \right]} \phi(m_i)$ with $\phi(m_i) = n_i$. 

Analogously, the canonical embedding allows us to work with the Gaussian distribution $D_{\bm{r}}$ with $\bm{r} \in {(\mathbb{R}^+)}^n$ over $\bigotimes_i H_i$ as a distribution over $\bigotimes_i K_i$. Actually, the distribution $D_{\bm{r}}$ is over $K_{(T),\mathbb{R}} = K_{(T)} \bigotimes_{\mathbb{Q}} \mathbb{R}$ which is also isomorphic to $H_{(T)}$ as a real vector space.\footnote{We will use $K_{(T)}$ instead of $K_{(T),\mathbb{R}}$ unless the distinction is relevant.} However, it is more helpful to ignore the distinction between $K_{(T)}$ and $K_{(T),\mathbb{R}}$ and to approximate the latter by the former using enough precision (in order to represent real numbers with rational numbers).

\subsubsection{Trace and Norm}

Here we present the basic concepts of trace and norm over number fields that were proposed in previous works. Section~\ref{specialproperties} highlights which are the changes needed and how we can work with them when we have the tensor product of non coprime cyclotomic fields.

The trace $\mbox{Tr} = \mbox{Tr}_{K/\mathbb{Q}}: K \rightarrow \mathbb{Q}$ and norm $N = N_{K/\mathbb{Q}}: K \rightarrow \mathbb{Q}$ are defined as:

\begin{equation}
\begin{split}
\mbox{Tr}(x) = \sum_{i \in [ n ]} \sigma_i(x), \mbox{ }&  N(x) = \prod_{i \in [ n ]} \sigma_i(x).
\end{split}
\end{equation}

In addition, the trace is a linear function in $\mathbb{Q}$ because $\mbox{Tr}(a + b) = \mbox{Tr}(a) + \mbox{Tr}(b)$ and $\mbox{Tr}(ca) = c\mbox{Tr}(a)$ for all $a,b \in K$ and $c \in \mathbb{Q}$. It is also important to note that $\mbox{Tr}(a \cdot b) = \sum_i \sigma_i(a)\sigma_i(b)$.

Even though we will do more emphasis later, we note that when working with tensor products $K_{(T)} = \bigotimes_i K_i$, resorting to the fact that $\sigma(\otimes_i a_i) = \otimes_i\sigma^{(i)}(a_i)$ the corresponding trace satisfies $\mbox{Tr}_{K_{(T)}/\mathbb{Q}}(\otimes_i a_i) = \prod_i \mbox{Tr}_{K_i/\mathbb{Q}}(a_i)$.

\subsubsection{Tensor Ring of Integers and its Ideals}

This appendix revises some basic properties of the ring of integers of a number field and its ideals. Although we are considering cyclotomic number fields $K_i = \mathbb{Q}(\varsigma_{m_i})$, these results apply to more general number fields. The ring of integers of a number field is denoted $\mathcal{O}_{K_i}$ and it is defined as the set of elements belonging to $K_i$ that satisfy a monic polynomial $f(x)$ with coefficients belonging to the integers, that is, elements $a \in K_i$ such that $f(a) = 0$.

It can be seen  that $\mathcal{O}_{K_i}$ is a free $\mathbb{Z}$-module with rank the degree of $K_i$ (when working with cyclotomic fields this degree is $\phi(m_i)$), and that its $\mathbb{Z}$-basis $B_i = \{ b_1^{(i)}, \ldots, b_n^{(i)} \} \subset \mathcal{O}_{K_i}$ results to be a $\mathbb{Q}$-basis for $K_i$ and also a $\mathbb{R}$-basis for ${K_i} \bigotimes \mathbb{R}$.

We work with the result of the tensor product of the different rings of integers which corresponds to each number field, that is, for the tensor of number fields $K_{(T)} = \bigotimes_{i \in \left[ l \right]} K_i$ we consider the tensor ring of integers $R = \bigotimes_{i \in \left[ l \right]} \mathcal{O}_{K_i}$. All the properties introduced for the ring of integers in \cite{LPR13} are also valid when working with ideals of the new multivariate polynomial ring $R$.

Firstly, we could see $R$ as a $\mathbb{Z}$-module with rank $n = \prod_{i \in \left[ l \right]} \phi(m_i)$ and its $\mathbb{Z}$-basis would be $\bigotimes_{i \in \left[ l \right]}B_i \subset R$ that also results to be a $\mathbb{Q}$-basis for $K_{(T)}$ and a $\mathbb{R}$-basis for $K_{(T), \mathbb{R}}$.

Next, we include some important facts about the ideals of $R$. An integral ideal (a.k.a. ideal) of $R$ is an additive subgroup that is closed under multiplication by $R$, that is, $r \cdot x \in \mathcal{I}$ for any $r \in R$ and $x \in \mathcal{I}$. In order to generate an ideal $\mathcal{I}$ of $R$, it can be shown that there exist two different elements $g_1, g_2 \in R$ whose $R$-linear combinations generate  $\mathcal{I} = \langle g_1, g_2 \rangle$. An ideal is also a free $\mathbb{Z}$-module of rank $n$, so we have some basis $\{ u_1, \ldots, u_n \} \subset R$.

The norm of an ideal is its corresponding index as an additive subgroup, that is, $N(\mathcal{I}) = |R : \mathcal{I}|$. The sum $\mathcal{I} + \mathcal{J}$ is also an ideal whose elements are all the pairs $x + y$ with $x \in \mathcal{I}$ and $y \in \mathcal{J}$, the product ideal $\mathcal{I}\mathcal{J}$ is the set of all finite sums of pairs $xy$ with $x \in \mathcal{I}$ and $y \in \mathcal{J}$. The norm of ideals generalizes the previous definition of norm in the following way $N(\langle x \rangle) = |N(x)|$ with $x \in R$ and $N(\mathcal{I}\mathcal{J}) = N(\mathcal{I})N(\mathcal{J})$.

We say that two ideals $\mathcal{I}$ and $\mathcal{J}$ are coprime (or relatively prime) if $\mathcal{I} + \mathcal{J} = R$. An ideal $\mathfrak{p} \subsetneq R$ is prime if whenever $ab \in \mathfrak{p}$ for some $a, b \in R$, then $a \in \mathfrak{p}$ or $b \in \mathfrak{p}$. An ideal $\mathfrak{p}$ of $R$ is prime if and only if it is maximal. The ring $R$ has unique factorization on ideals, that is, every ideal of $R$ can be expressed as a unique product of powers of prime ideals.

A fractional ideal $\mathcal{I} \subset K$ satisfies $d\mathcal{I} \subseteq R$ where $d\mathcal{I}$ is an integral ideal for some $d \in R$. Its norm is defined as $N(\mathcal{I}) = N(d\mathcal{I})/|N(d)|$.

\subsubsection{Ideal Lattices}

This work relies on the lattices embedded by the fractional ideals in $K_{(T)}$ under the canonical embedding. Next, we describe some of their properties. A fractional ideal $\mathcal{I}$ has a $\mathbb{Z}$-basis $U = \{ u_1, \ldots, u_n \}$. Then, under the canonical embedding $\sigma$, the ideal yields a rank-$n$ ideal lattice $\sigma(\mathcal{I})$ with basis $\{ \sigma(u_1), \ldots, \sigma(u_n) \} \subset H_{(T)}$. The lattice embedded by an ideal is commonly identified by the ideal, so we consider the minimum distance $\lambda_1(\mathcal{I})$ of an ideal.

The absolute discriminant $\Delta_{K}$ is defined for a field $K$. We generalize this term to the tensor field $K_{(T)}$, considering $\Delta_{K_{(T)}}$ as the square of the fundamental volume of the embedded lattice $\sigma(R)$. We also have $\Delta_{K_{(T)}} = |\mbox{det}(\mbox{Tr}(b_i \cdot b_j))|$, where $\{ b_1, \ldots, b_n \}$ is an integral basis of $R$. Therefore, we can define the fundamental volume of an ideal lattice $\sigma(\mathcal{I})$ as $N(\mathcal{I}) \cdot \sqrt{\Delta_{K_{(T)}}}$.

Now we include an important lemma that gives upper and lower bounds on the minimum distance of an ideal lattice.

\begin{lem}[Extended version of Lyubashevsky \emph{et al.} \cite{LPR13} Lemma $2.9$, Peikert and Rosen \cite{PR07} detailed proof\label{lem:2.9}] For any fractional ideal $\mathcal{I}$ in a tensor field $K_{(T)}$ of degree $n$, and in any $l_p$-norm for $p \in [ 1, \infty]$,
\begin{equation}
n^{1/p} \cdot {N(\mathcal{I})}^{1/n} \overset{(a)}{\leq} \lambda_1(\mathcal{I}) \overset{(b)}{\leq} n^{1/p} \cdot {N(\mathcal{I})}^{1/n} \cdot \sqrt{\Delta_{K_{(T)}}^{1/n}}.
\end{equation}
\end{lem}
The proof of the previous Lemma \ref{lem:2.9} follows analogously to the proofs of the Lemmas $6.1$ (upper bound) and $6.2$ (lower bound) in \cite{PR07}.

First, we start with the upper bound $(b)$ following the guidelines of \cite{PR07}. Considering ${||x||}_p \leq n^{1/p}{||x||}_{\infty}$ for $x \in K_{(T)}$, we only need to prove the bound for the $p = \infty$ norm. For this purpose, we resort to Minkowski's Theorem \ref{prop:4.1} to bound the distance of $\lambda_1^{\infty}$:

\begin{thmRLWE}[Minkowski's Theorem\label{prop:4.1}] Let $\Lambda$ be any lattice of rank $n$ and $\mathcal{B} \subseteq \mbox{span}\left( \Lambda \right)$ be any convex body symmetric about the origin having $n$-dimensional volume $\mbox{vol} \left( \mathcal{B} \right) > 2^n \cdot \mbox{det} \left( \Lambda \right)$. Then $\mathcal{B}$ contains some nonzero $\bm{x} \in \Lambda$.
\end{thmRLWE}

Now, we consider the $n$-dimensional closed $\mathcal{C} = \{ \bm{x} \in H_{(T)} : {||\bm{x}||}_{\infty} \leq 1 \}$, and each $\phi(m_i)$-dimensional closed $\mathcal{C}^{(i)} = \{ \bm{x} \in H_i : {||\bm{x}||}_{\infty} \leq 1 \}$. Knowing that $H_i \subseteq \mathbb{R}^{s_1^{(i)}} \times \mathbb{C}^{2s_2^{(i)}}$, it can be shown that the volume of $\mathcal{C}^{(i)}$ is $2^{\phi(m_i)} \cdot {(\pi/2)}^{s_2^{(i)}}$, where $\phi(m_i) = s_1^{(i)} + s_2^{(i)}$ and finally being $2^n \cdot {( \pi/2 )}^{\prod_{i \in \left[ l \right]} s_2^{(i)}}$ the volume of $\mathcal{C}$.

Proceeding as in \cite{PR07}, we have for any $\beta > N^{1/n} \left( \mathcal{I} \right) \cdot \sqrt{\Delta_{K_{(T)}}^{1/n}} \cdot {\left( 2/\pi \right)}^{\prod_{i \in \left[ l \right]} s_2^{(i)}/n}$
\begin{equation*}
\mbox{vol} \left( \beta \mathcal{C} \right) = \beta^n \mbox{vol} \left( \mathcal{C} \right) > 2^n \cdot N \left( \mathcal{I} \right) \cdot \sqrt{\Delta_{K_{(T)}}} = 2^n \cdot \mbox{det} \left( \sigma \left( \mathcal{I} \right) \right),
\end{equation*}
where by Minkowski's Theorem \ref{prop:4.1}, we know that $\beta \mathcal{C}$ contains a nonzero point of $\sigma \left( \mathcal{I} \right)$, therefore $\lambda_{1}^{\infty} \leq \beta$; consequently, it also satisfies the upper bound $(b)$ of Lemma \ref{lem:2.9}.

Regarding the lower bound $(a)$, we follow the steps of the proof for Lemma $6.2$ in \cite{PR07}. For $1 \leq p \leq \infty$, by the arithmetic mean/geometric mean inequality we have:
\begin{equation*}
{||x||}_p^p = \sum_{i \in \left[ n \right]} {|\sigma_i \left( x \right)|}^p \geq n \cdot {\left( \prod_{i \in \left[ n \right]} {|\sigma_i \left( x \right)|}^p \right)}^{1/n} = n \cdot {|N \left( x \right)|}^{p/n},
\end{equation*}
where by applying the $p$-root in both sides, it yields the considered lower bound $(a)$ by considering that $|N(x)| \geq N \left( \mathcal{I} \right)$ for any nonzero $x \in \mathcal{I}$ (for more details of both proofs we refer the reader to \cite{PR07}). Here, it is important to note that resorting to the concepts presented in Appendix~\ref{sec:embandgeom}, we can deal with the different embeddings, even when we are working with the tensor of number fields.

\subsubsection{Duality}

For any lattice $\mathcal{L}$ in $K_{(T)}$ (this is the $\mathbb{Z}$-span of any $\mathbb{Q}$-basis of $K_{(T)}$), its dual is defined as:
\begin{equation}
\mathcal{L}^{\vee} = \{ x \in K_{(T)} : \mbox{Tr}(x\mathcal{L}) \subseteq \mathbb{Z} \}.
\end{equation}

As in the ``traditional'' (non-tensor) number field case, using the canonical embedding, $\mathcal{L}^{\vee}$ embeds as the complex conjugate of the dual lattice, that is, $\sigma(\mathcal{L}^{\vee}) = \bar{\sigma}_{\mathcal{L}}^*$. Taking this into account and considering also that $\mathcal{L} = \bigotimes_{i \in \left[ l \right]} \mathcal{L}_i$ and the dual operation commutes the tensoring, we have:
\begin{align*}
  \sigma(\mathcal{L}^{\vee}) &= \sigma(\otimes_i \mathcal{L}_i^{\vee}) = \otimes_i \sigma(\mathcal{L}_i^{\vee}) = \otimes_i \bar{\sigma}^*(\mathcal{L}_i)\\
  &=\overline{ \otimes_i \sigma^*(\mathcal{L}_i)} = \overline{ {\left(\otimes_i \sigma(\mathcal{L}_i) \right)}^*} = \overline{ {\sigma}^*(\otimes_i \mathcal{L}_i)} = \overline{ {\sigma}^*(\mathcal{L})}.
\end{align*}

It is also easy to check that ${\left(\mathcal{L}^\vee \right)}^{\vee} = \mathcal{L}$ (tensoring commutes dual), and that if $\mathcal{L}$ is a fractional ideal, its dual is also fractional. 
An important fact is that an ideal and its inverse are related by multiplication with the dual ideal of the ring: for any fractional ideal $\mathcal{I}$, its dual ideal is $\mathcal{I}^{\vee} = \mathcal{I}^{-1} \cdot R^{\vee}$. The factor $R^{\vee}$ (often called codifferent) is a fractional ideal whose inverse ${(R^{\vee})}^{-1}$, called the different ideal, is integral and of norm $N({(R^{\vee})}^{-1}) = \Delta_{K_{(T)}}$, the discriminant of $K_{(T)}$.

\subsubsection{Ideal Lattice Problems}
\label{sec:idealLatticeProblems}
We revise here the computational problems over ideal lattices related to RLWE, and, by extension, to $m$-RLWE: the Shortest Vector Problem (SVP), Shortest Independent Vectors Problem (SIVP), and the Bounded Distance Decoding (BDD) Problem. The three problems can be restricted to the case of integral ideals over $R$ (the tensor of ring of integers $\mathcal{O}_{K_i}$), analogously to the argument followed by Lyubashevsky \emph{et al.} \cite{LPR10}, \cite{LPR13} in the non-tensor case: if $\mathcal{I}$ is a fractional ideal with denominator $d \in R$ (such that $d\mathcal{I} \subseteq R$ is a integral ideal), then the ideal $N(d) \cdot \mathcal{I} \subseteq R$, because $N(d) \in \langle d \rangle$.

\begin{defin}[SVP and SIVP] Let $K_{(T)}$ be a tensor of number fields endowed with some geometric norm (e.g, the $l_2$-norm), and let $\gamma \geq 1$. The $\mbox{K}_{(T)}\mbox{-SVP}_{\gamma}$ problem in the given norm is posed as: given a fractional ideal $\mathcal{I}$ in $K_{(T)}$, find some nonzero $x \in \mathcal{I}$ such that $||x|| \leq \gamma \cdot \lambda_1(\mathcal{I})$. The $\mbox{K}_{(T)}\mbox{-SIVP}_{\gamma}$ problem is defined similarly, where the goal is to find $n$ linearly independent elements in $\mathcal{I}$ whose norms are all at most $\gamma \cdot \lambda_n(\mathcal{I})$.
\end{defin}

\begin{defin}[BDD] Let $K_{(T)}$ be a tensor of number fields endowed with some geometric norm (e.g, the $l_2$ norm), let $\mathcal{I}$ be a fractional ideal in $K_{(T)}$, and let $d < \lambda_1(\mathcal{I})/2$. The $\mbox{K}_{(T)}\mbox{-BDD}_{\mathcal{I}, d}$ problem in the given norm is: given $\mathcal{I}$ and $y$ of the form $y = x + e$ for some $x \in \mathcal{I}$ and $||e|| \leq d$, find $x$.
\end{defin}

\subsubsection{Chinese Remainder Theorem}
\label{sec:CRT}

We reformulate the Chinese Remainder Theorem (CRT) for the ring $R = \bigotimes_{i \in \left[ l \right]} \mathcal{O}_{K_i}$ in the tensor of number fields $K_{(T)} = \bigotimes_{i \in \left[ l \right]} K_i$ and we also revisit some important concepts introduced in \cite{LPR13}.

\begin{lem}[Chinese Remainder Theorem\label{lem:CRT}] Let $\mathcal{I}_1, \ldots, \mathcal{I}_r$ be pairwise coprime ideals in $R$, and let $\mathcal{I} = \prod_{i \in [ r ]} \mathcal{I}_i$. The natural ring homomorphism $R \rightarrow \bigoplus_{i \in [ r ]} (R/\mathcal{I}_i)$ induces a ring isomorphism $R/\mathcal{I} \rightarrow \bigoplus_{i \in [ r ]} (R/\mathcal{I}_i)$.
\end{lem}

The next lemma states that when this ring isomorphism exists, we can compute a CRT basis $C$ for the set of pairwise coprime ideals $\mathcal{I}_1, \ldots, \mathcal{I}_r$. The basis is composed by elements $c_1, \ldots, c_r \in R$ that satisfy $c_i = 1 \bmod{ \mathcal{I}_i}$ and $c_i = 0 \bmod{\mathcal{I}_j}$ when $i \neq j$. We can use that basis in order to invert the CRT isomorphism as follows: for any $w = (w_1, \ldots, w_r) \in \bigoplus_i (R/\mathcal{I}_i)$, we have that $v = \sum_i w_i \cdot c_i \bmod{\mathcal{I}}$ is the unique element in $R/\mathcal{I}$ that maps to $w$ with that ring isomorphism.

\begin{lem}[Efficient computable basis for isomorphism] There is a deterministic polynomial-time algorithm that, given coprime ideals $\mathcal{I}, \mathcal{J} \subseteq R$ (represented by $\mathbb{Z}$-bases), outputs some $c \in \mathcal{J}$ such that $c = 1 \bmod{\mathcal{I}}$. More generally, there is a deterministic polynomial-time algorithm that, given pairwise coprime ideals $\mathcal{I}_1, \ldots, \mathcal{I}_r$, outputs a CRT basis $c_1, \ldots, c_r \in R$ for those ideals.
\end{lem}

Now we include two more lemmas that allow us to efficiently compute a bijection between the quotient groups $\mathcal{I}/q\mathcal{I}$ and $\mathcal{J}/q\mathcal{J}$ for any fractional ideals $\mathcal{I}, \mathcal{J}$. They are important for clearing out the arbitrary ideal $\mathcal{I}$ in the BDD-to-LWE reduction. The lemmas are:

\begin{lem}[Lyubashevsky \emph{et al.} \cite{LPR13} Lemma $2.14$\label{lem:2.14}] Let $\mathcal{I}$ and $\mathcal{J}$ be ideals in $R$. There exists $t \in \mathcal{I}$ such that the ideal $t \cdot \mathcal{I}^{-1} \subseteq R$ is coprime to $\mathcal{J}$. Moreover, such $t$ can be found efficiently given $\mathcal{I}$ and the prime ideal factorization of $\mathcal{J}$.
\end{lem}

\begin{lem}[Lyubashevsky \emph{et al.} \cite{LPR13} Lemma $2.15$\label{lem:2.15}] Let $\mathcal{I}$ and $\mathcal{J}$ be ideals in $R$, let $t \in \mathcal{I}$ be such that $t \cdot \mathcal{I}^{-1}$ is coprime with $\mathcal{J}$, and let $\mathcal{M}$ be any fractional ideal in $K_{(T)}$. Then, the function $\theta_t: K_{(T)} \rightarrow K_{(T)}$ defined as $\theta_t(u) = t \cdot u$ induces an isomorphism from $\mathcal{M}/\mathcal{J}\mathcal{M}$ to $\mathcal{I}\mathcal{M}/\mathcal{I}\mathcal{J}\mathcal{M}$, as $R$-modules. Moreover, this isomorphism may be efficiently inverted given $\mathcal{I}$, $\mathcal{J}$, $\mathcal{M}$ and $t$.
\end{lem}
The proof of Lemma \ref{lem:2.15} for the case where $K_{(T)}$ is a tensor of cylotomic fields follows with the same techniques considered in \cite{LPR13}, by taking into account that $\theta_t$ induces a homomorphism of $R$-modules because it represents a multiplication by a $t \in R$, so we do not include it here.

\section{Proof of Theorem \ref{th:hardsearchLWE}}
\label{sec:proofhardsearchLWE}
This appendix presents the proof of Theorem \ref{th:hardsearchLWE}. It is based on the iterative use of the following lemma:

\begin{lem}[Extended version of Lemma $4.2$ Lyubashevsky \emph{et al.} \cite{LPR13}\label{lem:proofthhardsearchLWE}] Let $\alpha > 0$ and $q \geq 2$ be an integer. There exists an efficient quantum algorithm that, given a fractional ideal $\mathcal{I}$ in $K_{(T)}$, a number $r \geq \sqrt{2}q \cdot \eta_{\epsilon}(\mathcal{I})$ for some negligible $\epsilon = \epsilon(n)$ such that $r' = r \cdot \omega(\sqrt{\log{n}})/(\alpha q) > \sqrt{2n}/\lambda_1(\mathcal{I}^{\vee})$, an oracle to $m\mbox{-R-LWE}_{q, \Psi_{\leq \alpha}}$, and a list of samples from the discrete Gaussian distribution $D_{\mathcal{I}, r}$ (as many as required by the $m\mbox{-R-LWE}_{q, \Psi_{\leq \alpha}}$ oracle), outputs an independent sample from $D_{\mathcal{I}, r'}$.
\end{lem}

Theorem \ref{th:hardsearchLWE} is proven as follows: we start with a value $r \geq 2^{2n}\lambda_n(\mathcal{I})$, in such a way that we can classically generate any polynomial number of samples from $D_{\mathcal{I}, r}$. Given the samples from $D_{\mathcal{I}, r}$, Lemma \ref{lem:proofthhardsearchLWE} can be used iteratively a polynomial number of times (using the same samples) to obtain a polynomial number of independent samples from $D_{\mathcal{I}, r'}$ with $r' = r/2$ at each iteration. Repeating this process, we can obtain samples from narrower and narrower distributions, until we have samples from a distribution with parameter $s \geq \gamma$.

Lemma \ref{lem:proofthhardsearchLWE} is obtained thanks to the following two results (Lemmas \ref{lem:4.3} and \ref{lem:4.4}):

\begin{lem}[Extended version of Lemma $4.3$ of Lyubashevsky \emph{et al.} \cite{LPR13}, proof in Section $4.2$\label{lem:4.3}] Let $\alpha > 0$, let $q \geq 2$ be an integer with known factorization, let $\mathcal{I}$ be a fractional ideal in $K_{(T)}$, and let $r \geq \sqrt{2}q \cdot \eta_{\epsilon}(\mathcal{I})$ for some negligible $\epsilon = \epsilon(n)$. Given an oracle for the discrete Gaussian distribution $D_{\mathcal{I}, r}$, there is a probabilistic polynomial-time (classical) reduction from $\mbox{BDD}_{\mathcal{I}^{\vee}, d}$ in the $l_{\infty}$ norm to $m\mbox{-R-LWE}_{q, \Psi_{\leq \alpha}}$, where $d = \alpha q/(\sqrt{2}r)$.
\end{lem}

Details for the proof of the lemma \ref{lem:4.3} follow the same steps of Lyubashevsky \emph{et al.} for Lemma $4.3$ in \cite{LPR13}, so we do not replicate it here. However, we have to take into account that we are working with ideals over the tensor of the ring of integers, so instead of considering the lemmas $2.14$ and $2.15$ from \cite{LPR13} we have to use the redefined lemmas already presented in our work as Lemmas \ref{lem:2.14} and \ref{lem:2.15}.

\begin{lem}[Extended version of Lemma $4.4$ of Lyubashevsky \emph{et al.} \cite{LPR13}\label{lem:4.4}] There is an efficient quantum algorithm that, given any $n$-dimensional lattice $\Lambda$, a number $d' < \lambda_1(\Lambda^{\vee})/2$ (where $\lambda_1$ is with respect to the $l_2$ norm), and an oracle that solves BDD on $\Lambda^{\vee}$ except with negligible probability for points whose offset from $\Lambda^{\vee}$ is sampled from $D_{d'/\sqrt{2n}}$, outputs a sample from $D_{\Lambda, \sqrt{n}/(\sqrt{2}d')}$. In particular, since a sample from $D_{d'/\sqrt{2n}}$ has $l_{\infty}$ norm at most $d' \cdot \omega(\sqrt{\log{n}})/\sqrt{n}$ except with negligible probability, it suffices if the oracle solves $\mbox{BDD}_{\mathcal{I}^{\vee}, d}$ in the $l_{\infty}$ norm, where $d = d' \cdot \omega(\sqrt{\log{n}})/\sqrt{n}$.
\end{lem}

The sketch of the proof for the lemma \ref{lem:proofthhardsearchLWE} is the following: starting with samples from $D_{\mathcal{I}, r}$ and an oracle for $m\mbox{-R-LWE}_{q, \Psi_{\leq \alpha}}$ and resorting to the lemma \ref{lem:4.3} we can obtain an algorithm for BDD on $\mathcal{I}^{\vee}$ to within distance $d = \alpha q/(\sqrt{2}r)$ in the $l_{\infty}$ norm. Next, considering Lemma \ref{lem:4.4} with $d' = d \sqrt{n} / \omega(\sqrt{\log{n}}) = \sqrt{n/2}/r' < \lambda_1(\mathcal{I}^{\vee})/2$, we obtain a quantum procedure that outputs samples from the discrete Gaussian distribution $D_{\mathcal{I}, r'}$.

\section{Proofs of Theorems~\ref{th:pseudorandomness1}, \ref{th:pseudorandomness2} and \ref{th:pseudorandomness3}}
\label{sec:proofs4everyting}
This appendix includes the proofs for the main results involving the security reductions of $m$-RLWE, as stated in Theorems~\ref{th:pseudorandomness1}, \ref{th:pseudorandomness2} and \ref{th:pseudorandomness3}.

\subsection{Search to Worst-Case Decision}

Here we explain the two first reductions of the Theorems \ref{th:pseudorandomness1} and \ref{th:pseudorandomness2}. Next, we introduce the main definitions of the intermediate problems and the corresponding lemmas, and we also highlight the differences due to working with the tensor of the rings of integers.

\begin{defin}[Extended version of the $\mathfrak{q}_i\mbox{-LWE}_{q, \Psi}$ problem, Definition $5.4$ from Lyubashevsky \emph{et al.} \cite{LPR13}] The $\mathfrak{q}_i\mbox{-LWE}_{q, \Psi}$ problem is defined as: given access to $A_{s, \psi}$ for some arbitrary $s \in R_q^{\vee}$ and $\psi \in \Psi$, find $s \bmod{\mathfrak{q}_iR^{\vee}}$.
\end{defin}

\begin{lem}[LWE to $\mathfrak{q}_i\mbox{-LWE}$, entending Lemma $5.5$ of Lyubashevsky \emph{et al.} \cite{LPR13}\label{lem:5.5}] Suppose that the famility $\Psi$ is closed under all the automorphisms of $K_{(T)}$ (see Lemma \ref{lem:5.6}), that is, $\psi \in \Psi$ implies that $\uptau_k(\psi) \in \Psi$ for all $k \in \left[n\right]$. Then, for every $i \in \left[ n \right]$, there exists a deterministic polynomial-time reduction from $\mbox{LWE}_{q, \Psi}$ to $\mathfrak{q}_i\mbox{-LWE}_{q, \Psi}$.
\end{lem}

The proof is based on the fact that by having an oracle for $\mathfrak{q}_i\mbox{-LWE}$ and resorting to the different field automorphisms, we can recover $s$ modulo $\mathfrak{q}_jR^{\vee}$ for every $j \in \left[ n \right]$ and we can use the CRT for recovering $s$ modulo $R^{\vee}$.

The reduction works in the following way: Let $(a, b) \leftarrow A_{s, \psi}$ and apply an automorphism $(\uptau_k(a), \uptau_k(b))$ that satisfies $\uptau_k(\mathfrak{q}_j) = \mathfrak{q}_{i}$. Now, we use the $\mathfrak{q}_i\mbox{-LWE}$ oracle with the transformed samples and we apply the reverse automorphism ${\uptau_{k}(t)}^{-1} \in R^{\vee}/\mathfrak{q}_jR^{\vee}$ to its output $t \in R^{\vee}/\mathfrak{q}_iR^{\vee}$.

In order to see that ${\uptau_k(t)}^{-1}$ has the desired value $s \bmod{\mathfrak{q}_jR^{\vee}}$, we operate with the pair $(\uptau_k(a), \uptau_k(b))$, with $\uptau_k(b) = \uptau_k(a) \cdot \uptau_k(s)/q + \uptau_k(e) \bmod{R^{\vee}}$ where we see that the pair follows the $A_{\uptau_k(s), \uptau_k(\psi)}$ distribution (we know that $\uptau_k(\psi) \in \Psi$, see Lemma \ref{lem:5.6}). Therefore, the oracle outputs $t = \uptau_k(s) \bmod{\mathfrak{q}_iR^{\vee}}$ and Lemma \ref{lem:5.5} is proven.

\begin{lem}[Extended version of Lemma $5.6$ of Lyubashevsky \emph{et al.} \cite{LPR13}\label{lem:5.6}] For any $\alpha > 0$, the family $\Psi_{\leq \alpha}$ is closed under every automorphism $\uptau$ of $K_{(T)}$, that is, $\psi \in \Psi_{\leq \alpha}$ implies that $\uptau(\psi) \in \Psi_{\leq \alpha}$.
\end{lem}

In order to see that for $\psi \in \Psi$ any possible automorphism also belongs to $\Psi$, we proceed as follows: each automorphism is the tensor of the existing automorphisms for each cyclotomic field, that is, $\otimes_{i \in \left[ l \right]} \uptau_{k_i}^{(i)}$ with $k_i \in \mathbb{Z}_{m_i}^*$. Hence, resorting to the definition of our error distributions (see Appendix~\ref{sec:gaussianmeasure}), we have $\psi = D_{\otimes_{i \in \left[ l \right]} \bm{r}_i} \in \Psi_{\leq \alpha}$ where the elements of $\otimes_{i \in \left[ l \right]}\bm{r}_i$ are bounded by $\alpha$. As the effect of the automorphism simply permutes the coordinates of each $\bm{r}_i$, we can clearly see that $\otimes_{j \in \left[ l \right]} \uptau_{k_j}^{(j)}\left( D_{\otimes_{i \in \left[ l \right]} \bm{r}_i} \right) = D_{\otimes_{i \in \left[ l \right]} \bm{r}_i'}$ for $k_j \in \mathbb{Z}_j^*$, which also belongs to $\Psi_{\leq \alpha}$ because the value of the different elements follow being at most $\alpha$ (they have only been permuted).

We now move on to Lemma \ref{lem:5.9} for the second reduction of the proof, but we first introduce two definitions for the intermediate problems:

\begin{defin}[Extended Hybrid LWE Distribution of Lyubashvesky \emph{et al.}\cite{LPR13}\label{def:5.7}] For $j \in \left[ n \right]$, $s \in R_q^{\vee}$, and a distribution $\psi$ over $K_{(T), \mathbb{R}}$, the distribution $A_{s, \psi}^{j}$ over $R_q \times \mathbb{T}$ is defined as follows: choose $(a, b) \leftarrow A_{s, \psi}$ and output $(a, b + h/q)$ where $h \in R_q^{\vee}$ is uniformly random and independent modulo $\mathfrak{q}_iR^{\vee}$ for all $i \leq j$, and is equal to zero modulo all the remaining $\mathfrak{q}_iR^{\vee}$. We also define $A_{s, \psi}^{0} = A_{s, \psi}$.
\end{defin}

\begin{defin}[Extended $\mbox{WDLWE}_{q, \Psi}^{j}$ (Worst-Case Decision LWE Relative to $\mathfrak{q}_j$) of Lyubashevsky \emph{et al.} \cite{LPR13}\label{def:5.8}] For $j \in \left[ n \right]$ and a family of distributions $\Psi$, the $\mbox{WDLWE}_{q, \Psi}^{j}$ problem is defined as follows: given access to $A_{s, \psi}^{i}$ for arbitrary $s \in R_q^{\vee}$, $\psi \in \Psi$, and $i \in \{j - 1, j \}$, find $i$.
\end{defin}

\begin{lem}[Extended version of Search to Decision of Lyubashvesky \emph{et al.} \cite{LPR13}\label{lem:5.9}] For any $j \in \left[ n \right]$, there exists a probabilistic polynomial-time reduction from $\mathfrak{q}_j\mbox{-LWE}_{q, \Psi}$ to $\mbox{WDLWE}_{q, \Psi}^{j}$.
\end{lem}

The proof of the reduction is based on trying each of the different possible values of $s$ modulo $\mathfrak{q}_jR^{\vee}$ in such a way that after modifying the samples from $A_{q, \psi}$, we have that a) for the correct value, the samples are distributed following $A_{q, \psi}^{j - 1}$ and b) for the rest of possible values, they follow $A_{q, \psi}^{j}$.

We can try all different values for $s \bmod{\mathfrak{q}_jR^{\vee}}$ because the norm of $\mathfrak{q}_j$ for all $j$ satisfies $N(\mathfrak{q}_j) = q = \mbox{poly}(n)$, so we can enumerate all the combinations. Finally, we can use the oracle $\mbox{WDLWE}_{q, \Psi}^{j}$ for distinguishing between the distributions $A_{q, \psi}^{j - 1}$ and $A_{q, \psi}^{j}$.

Following an analogous procedure as the one in \cite{LPR13}, given a sample $(a, b) \leftarrow A_{s, \psi}$, we have:
\begin{equation*}
(a', b') = (a + v, b + (h + vg)/q) \in R_q \times \mathbb{T},
\end{equation*}
where $v \in R_q$ satisfies that it is uniformly random modulo $\mathfrak{q}_j$ and zero modulo other different prime ideal, $h, g \in R_q^{\vee}$, where $h$ is uniformly random and independent modulo any $\mathfrak{q}_iR^{\vee}$ when $i < j$, and it is zero for the rest of possible values of $i$. Finally, we have:
\begin{equation*}
b' = (a's + h + v(g - s))/q + e,
\end{equation*}
with $e \leftarrow \psi$.

Now, choosing different values for $g$ we have the following results: a) if $g = s \bmod{\mathfrak{q}_jR^{\vee}}$, the distribution of $(a', b')$ is $A_{s, \psi}^{j - 1}$, and b) if $g \neq s \bmod{\mathfrak{q}_jR^{\vee}}$, the distribution of $(a', b')$ is $A_{s, \psi}^{j}$. Hence, we only have to enumerate different $g$ values which satisfy different conditions modulo $\mathfrak{q}_jR^{\vee}$ (the values modulo other $\mathfrak{q}_iR^{\vee}$ with $i \neq j$ are not important) to achieve the reduction.

\subsection{Worst-Case Decision to Average-Case Decision}

The objective of this part is to cover the two last reductions of Theorems \ref{th:pseudorandomness1} and \ref{th:pseudorandomness2}. For this purpose, we present some definitions and lemmas that allow us to reduce the worst-case decision $\mbox{WDLWE}_{q, \Psi}^{j}$ problem to an average-case problem $\mbox{DLWE}_{q, \Upsilon}$ where the goal is to distinguish between $A_{s, \psi}$ and uniform samples where the parameters of the error distribution are also secret and drawn from $\Upsilon$.

\begin{defin}[Extended version of Average-Case Decision LWE Relative to $\mathfrak{q}_j$ ($\mbox{DLWE}_{q, \Upsilon}^{j}$) of Lyubashevsky \emph{et al.} \cite{LPR13}\label{def:5.10}] For $j \in \left[ n \right]$ and a distribution $\Upsilon$ over error distributions, we say that an algorithm solves the $\mbox{DLWE}_{q, \Upsilon}^{j}$ problem if with a non negligible probability over the choice of a random $(s, \psi) \leftarrow U(R_q^{\vee}) \times \Upsilon$, it has a non negligible difference in acceptance probability on inputs from $A_{s, \psi}^{j}$ versus inputs from $A_{s, \psi}^{j - 1}$.
\end{defin}

\begin{lem}[Extended version of Worst-Case to Average-Case Lemma $5.12$ of Lyubashevsky \emph{et al.} \cite{LPR13}\label{lem:5.12}] For any $\alpha > 0$ and every $j \in \left[ n \right]$, there is a randomized polynomial-time reduction from $\mbox{WDLWE}_{1, \Psi_{\leq \alpha}}^{j}$ to $\mbox{DLWE}_{q, \Upsilon_{\alpha}}^{j}$.
\end{lem}

In order to prove the previous lemma, let $s' \in R_q^{\vee}$, $\bm{r}' \in { \left( \mathbb{R}^+ \right)}^n$, $k \in \left[ n \right]$, and the pair $(a, b)$, and consider the transformation $(a, b + (a \cdot s' + h)/q + e')$ where $e'$ is drawn from $D_{\bm{r}'}$, $h \in R_q^{\vee}$ and $h$ satisfies that $h \bmod{\mathfrak{q}_iR^{\vee}}$ are uniformly random and independent for $i \leq k$, and zero for all other $i$. Then, when the input is $A_{s, \psi}^{j}$, this transformation outputs $A_{s + s', \psi + D_{\bm{r}'}}^{\max{ \{ k, j \} }}$.

Now, to achieve the reduction, we repeat the following process a polynomial number of times: we draw $s' \in R_q^{\vee}$, and we have $\bm{r}' \in {\left( \mathbb{R}^+ \right)}^n$ where $\bm{r}' = \bigotimes_{i \in \left[ l \right]} \bm{r}_i'$ (as it was presented in Appendix~\ref{sec:gaussianmeasure}) and $r'_{i,j} = r'_{i, j + \phi(m_i)/2}$ with $i \in \left[ l \right]$ and $j \in \left[ \phi(m_i) \right]$. We also have ${r'}_j^2 = \alpha^2\sqrt{n}x_j$ and ${r'}_i^2 = \alpha^2\sqrt{n}x_i$ for all $j, i \in \left[ n \right]$ and where the $x_j$ and $x_i$ are chosen independently from $\Gamma(2,1)$ if $r_j$ and $r_i$ are different. Next, we estimate the acceptance probability of the oracle for two different input distributions: a) applying to the input the previous transformation with parameters $s'$, $\bm{r}'$ and $j-1$; b) applying to the input the previous transformation with parameters $s'$, $\bm{r}'$ and $j$. Finally, after a polynomial number of repetitions we output $j - 1$ if there is a non negligible difference between the two acceptance probabilities; on the contrary, we output $j$.

Let us assume that the input distribution is $A_{q, D_{\bm{r}}}^{j - 1}$ for some $\bm{r}$ where all $r_{i} \in \left[0, \alpha \right]$ for $i \in \left[ n \right]$. Then, we have to estimate the acceptance probability of the oracle on $A_{s + s', D_{\bm{r}} + D_{\bm{r}'}}^{j - 1}$ and $A_{s + s', D_{\bm{r}} + D_{\bm{r}'}}^{j}$, and we notice that $D_{\bm{r}} + D_{\bm{r}'} = D_{\bm{r}''}$ where ${r''}^2_{i} = {r'}^2_{i} + r^2_{i}$. If we denote by $S$ the set of pairs $(s, \psi)$ for which the oracle has non negligible difference in acceptance probability between $A_{q, \psi}^{j - 1}$ and $A_{q, \psi}^{j}$, we have by assumption (the measure of $S$ under $U(R_q^{\vee}) \times \Upsilon_{\alpha}$ is non negligible) and by claim \ref{claim:5.11} below that $(s + s', D_{\bm{r}} + D_{\bm{r}'}) \in S$ with non negligible probability, and the proof of Lemma \ref{lem:5.12} is complete.

Our Claim \ref{claim:5.11} a variant of the Claim $5.11$ presented by Lyubashevsky \emph{et al.} \cite{LPR13}. For our case, we need a similar result, but it must hold not only for independent variables following a $\Gamma(2, 1)$ distribution, because in our more general case, for $i \in \left[ n \right]$ we can have that more than two $x_i$ are equal. Therefore, we present a modification for vectors of coefficients distributed as $\Gamma(2, 1)$, where they do not have to be independent, and we justify its validity.

\begin{claim}[Extended Claim $5.11$ from \cite{LPR13}\label{claim:5.11}] Let $P$ be the distribution ${\Gamma(2,1)}^n$ and $Q$ be the distribution $\left(\Gamma(2, 1) - z_1 \right) \times \dots \times \left( \Gamma(2,1) - z_n \right)$ for some $0 \leq z_1, \ldots, z_n \leq 1/\sqrt{n}$ where the different $\Gamma(2, 1)$ of both $P$ and $Q$ do not have to be independent and some of them can be equal to each other. Then, any set $A \subseteq \mathbb{R}^n$ whose measure under P is non negligible also has non negligible measure under $Q$.
\end{claim}

The proof of the claim follows the next scheme: first, let $P, Q: \mathbb{R}^{n} \rightarrow \mathbb{R}^{+}$, where when $Q(\bm{x}) = 0$ we also have $P(\bm{x}) = 0$, and we define $R \left( P || Q \right) = \int_{\mathbb{R}^n}{\frac{{P(\bm{x})}^{2}}{Q(\bm{x})}d\bm{x}}$, considering that the fraction is zero when both the numerator and the denominator are zero. By Cauchy-Schwarz inequality, we have for any non empty set $A \subseteq \mathbb{R}^n$,
\begin{equation*}
\frac{{\left( \int_A{P(\bm{x}) d\bm{x}} \right)}^2}{\int_A{Q(\bm{x}) d\bm{x}}} \leq \int_{A}{\frac{{P(\bm{x})}^2}{Q(\bm{x})}d\bm{x}} \leq R \left( P || Q \right).
\end{equation*}
Thus, if we have a set $A$ with non negligible measure under $P$ and $R \left( P || Q \right) \leq \mbox{poly}(n)$ holds, we can say that the set $A$ has non negligible measure under $Q$.

For the particular setting of the Claim \ref{claim:5.11}, when $z > 0$ we have
\begin{equation*}
  R \left( \Gamma(2, 1) || \Gamma(2, 1) - z \right) = e^z \left( 1 - z + z^2 e^z \int_{z}^{\infty} x^{-1} e^{-x} dx\right),
\end{equation*}
and when $z$ is small, this expression reduces to $1 + z^2 \log{(1/z)} + \mathcal{O}(z^2)$.

The difference regarding the proof of \cite{LPR13} relies on the following fact: if we compute $R \left( P || Q \right)$, we have:
\begin{align*}
  &R \left( {\Gamma(2, 1)}^n || \left( \Gamma(2, 1) - z_1 \times \dots \times \Gamma(2, 1) - z_n \right) \right)\\
  &\leq R \left( \Gamma(2, 1) || \Gamma(2, 1) - z_1 \right) \dots R \left( \Gamma(2, 1) || \Gamma(2, 1) - z_n \right),
\end{align*}
where the equality is achieved when all the components of each vector are independent. When some of the $\Gamma(2, 1)$ variables are equal, we can see that the ratio of the corresponding distributions is equal to the ratio of only one of the variables of $P$ and $Q$ respectively.

Now, as we know that the second term of the expression is bounded by $\mbox{poly}(n)$, the claim is proven because for the setting of the claim our expression is bounded by the second term.

\begin{lem}[Extended version of Lemma $5.14$ Hybrid by Lyubashevsky \emph{et al.} \cite{LPR13}\label{lem:5.14}] Let $\Upsilon$ be a distribution over noise distributions satisfying that for any $\psi$ in the support of $\Upsilon$ and any $s \in R_q^{\vee}$, the distribution $A_{s, \psi}^{n}$ is within negligible statistical distance from uniform. Then for any oracle solving the $\mbox{DLWE}_{q, \Upsilon}$ problem, there exists a $j \in \left[ n \right]$ and an efficient algorithm that solves $\mbox{DLWE}_{q, \Upsilon}^{j}$ using the oracle.
\end{lem}

The proof works as follows: consider a pair $(s, \psi)$ for which the oracle can distinguish between $A_{s, \psi}$ and uniform distribution with a non negligible advantage. By Markov's inequality, the probability measure of those pairs is non negligible. Knowing that $A_{s, \psi}^{0} = A_{s, \psi}$ and that $A_{s, \psi}^{n}$ is negligibly far from the uniform distribution (see Lemma \ref{lem:5.13}), we see that for each $(s, \psi)$ we must have  a $j \in \left[ n \right]$ for which the oracle distinguishes between $A_{q, \psi}^j$ and $A_{q, \psi}^{j - 1}$ with non negligible advantage. Finally, the lemma is proven if we take the $j$ that is associated to the set of pairs $(s, \psi)$ with the highest probability. With the proof of this lemma, the proof of the Theorem \ref{th:pseudorandomness1} is complete.

\begin{lem}[Adapted version of lemma $5.13$ of Lyubashevsky \emph{et al.} \cite{LPR13}\label{lem:5.13}] Let $\alpha \geq \eta_{\epsilon}(R^{\vee})/q$ for some $\epsilon > 0$. Then, for any $\psi$ in the support of $\Upsilon_{\alpha}$ and $s \in R_q^{\vee}$, the distribution $A_{s, \psi}^{n}$ is within statistical distance $\epsilon/2$ of the uniform distribution over $(R_q, \mathbb{T})$.
\end{lem}
The proof of this lemma is obtained by following the steps in \cite{LPR13} and taking into account the considered changes in our setting together with our Lemma \ref{lem:2.3}.

Finally, we introduce the needed lemma for the reductions of Theorem \ref{th:pseudorandomness2}.

\begin{lem}[Extended version of Lemma $5.16$ of Lyubashevsky \emph{et al.} \cite{LPR13} Worst-Case to Average-Case with Spherical Noise\label{lem:5.16}] For any $\alpha > 0$, $l \geq 1$, and every $j \in \left[ n \right]$, there exists a randomized polynomial-time reduction from solving $\mbox{WDLWE}_{q, \Psi_{\leq \alpha}}^{j}$ to solving $\mbox{DLWE}_{q, D_{\xi}}^{j}$ given only $l$ samples, where $\xi = \alpha{\left( nl/\log{(nl)} \right)}^{1/4}$.
\end{lem}

In order to prove the Lemma \ref{lem:5.16}, we consider the transformation that we have already used for the proof of the Lemma \ref{lem:5.12}, but in this case the transformation has $l$ different inputs. So, let $s' \in R_q^{\vee}$, $k \in \left[ n \right]$, and $e_i \in \mathbb{T}$ for $i \in \left[ l \right]$. Now, consider for the following $l$ samples $(a_i, b_i)$ the mentioned transformation $(a_i, b_i + (a_i \cdot s' + h_i)/q + e_i)$, where $h_i \in R_q^{\vee}$ and $i \in \left[ l \right]$. It is important to note that all the $h_i$ satisfy that they are independent and uniform modulo $\mathfrak{q}_dR^{\vee}$ for all $d \leq k$, and they are zero when $d$ does not satisfy the previous relation. Therefore, if we take $l$ independent inputs drawn from $A_{q, \psi}^{j}$ and we apply the transformation to all of them considering that all $e_i$ are independently drawn from $D_{\bm{r}'}$, we have as output distribution ${\left( A_{s + s', \psi + D_{\bm{r}'}}^{\max{\{ k, j \}}} \right)}^l$.

Now, the reduction repeats the following process a polynomial number of times: we consider $s' \in R_q^{\vee}$ and a set of independent $e_i$ drawn from $D_{\xi}$. Next, we estimate the acceptance probability of the oracle for two different input distributions: a) applying to the input the previous transformation with parameters $s'$, $e_i$ and $j - 1$; b) applying to the input the previous transformation with parameters $s'$, $e_i$ and $j$. After a polynomial number of repetitions, we output $j - 1$ whenever a non negligible difference between the two acceptance probabilities is observed; otherwise, we output $j$.

Assuming the input distribution is $A_{s, D_{\bm{r}}}^{j - 1}$, where all the coefficients of $\bm{r}$ are in $\left[ 0, \alpha \right]$ for the two previous cases, we have two different output distributions: ${\left( A_{s + s', \psi + D_{\bm{r}'}}^{j - 1} \right)}^l$ and ${\left( A_{s + s', \psi + D_{\bm{r}'}}^{j}\right)}^l$. We also consider that the coefficients of $\bm{r}'$ verify ${r'}_i^2 = \xi^2 - r_i^2$, so we have $D_{\bm{r}} + D_{\bm{r}'} = D_{\xi}$.

As with Lemma \ref{lem:5.12}, let $S$ be the set of all tuples $(s, e_1, \ldots, e_l)$ for which the oracle has a non negligible difference in acceptance probability on ${\left( A_{s + s', \psi + D_{\bm{r}'}}^{j - 1} \right)}^l$ and ${\left( A_{s + s', \psi + D_{\bm{r}'}}^{j}\right)}^l$. By our assumption and a Markov argument, the measure of $S$ under $U \left( R_q^{\vee} \right) \times {\left( D_{\bm{r}'} \right)}^l$ is non negligible, and we have
\begin{equation*}
1 \leq \frac{\xi}{\sqrt{\xi^2 - r_i^2}} \leq \frac{\xi}{\sqrt{\xi^2 - \alpha^2}} \leq 1 + \sqrt{\frac{\log{(nl)}}{nl}},
\end{equation*}
where thanks to the Claim \ref{claim:5.15} below, we can assert that $S$ is also non negligible under $U \left( R_q^{\vee} \right) \times {\left( D_{\xi} \right)}^l$, and where we can derive the condition $\xi = \alpha{\left( nl/\log{(nl)} \right)}^{1/4}$, hence completing the proof of the Lemma \ref{lem:5.16} and the Theorem \ref{th:pseudorandomness2}.

\begin{claim}[Claim $5.15$ from \cite{LPR13}\label{claim:5.15}] Let $r_1, \ldots, r_n \in \mathbb{R}^+$ and $s_1, \ldots, s_n \in \mathbb{R}^+$ be such that for all $i$, $|s_i/r_i - 1| < \sqrt{\left(\log{n}\right)/n}$. Then any set $A \subseteq \mathbb{R}^n$ whose measure under the Gaussian distribution $D_{r_1} \times \dots \times D_{r_n}$ is non negligible, also has non negligible measure under $D_{s_1} \times \dots \times D_{s_n}$.
\end{claim}

\end{document}